\documentclass[twocolumn]{aastex7}
\usepackage{graphicx}
\usepackage{epstopdf}
\usepackage{natbib}
\usepackage[hyphens]{url}
\usepackage{amsmath}

\newcommand{\ee}[1]{\mbox{${} \times 10^{#1}$}}% scientific number format
\newcommand{\eten}[1]{\mbox{$10^{#1}$}}% power of ten

\newcommand{\kms}{\mbox{km s$^{-1}$}}% km/s
\newcommand\cmv{\mbox{cm$^{-3}$}}

\newcommand{\x}{\mbox{${}\times{}$}}
\newcommand{\sfr}{\mbox{$\rm{SFR}$}}

\newcommand{\sfrseventy}{\mbox{${\rm SFR}_{\rm 70}$}}

\newcommand{\sigmasfr}{\mbox{$\Sigma_{\rm SFR}$}}
\newcommand{\sigmastar}{\mbox{$\Sigma_{\star}$}}
\newcommand{\sigmam}{\mbox{$\Sigma_{\rm mol}$}}
\newcommand{\masscloud}{\mbox{$M_{\rm cloud}$}}
\newcommand{\areacloud}{\mbox{$A_{\rm cloud}$}}
\newcommand{\massp}{\mbox{$M^{\rm beam}_{\rm mol}$}}
\newcommand{\masscell}{\mbox{$M_{\rm cell}$}}
\newcommand{\sigmamcloud}{\mbox{$\Sigma_{\rm cloud}$}}
\newcommand{\sigmolcloud}{\mbox{$\Sigma^{\rm cloud}_{\rm mol}$}} % AL25 
\newcommand{\sigmamcell}{\mbox{$\Sigma_{\rm cell}$}}
\newcommand{\sigmamcloudmean}{\mbox{$\mean{\Sigma_{\rm cloud}}$}}
\newcommand{\sigmacloudhex}{\mbox{$\mean{\Sigma_{\rm cloud}}_{\rm hex}$}}
\newcommand{\sigsmhex}{\mbox{$\mean{\Sigma_{\rm cell}}_{\rm hex}$}}
\newcommand{\sigmav}{\mbox{$\sigma_{\rm v}$}}
\newcommand{\sigvcloud}{\mbox{$\sigma_{\rm v}^{\rm cloud}$}}
\newcommand{\sigvp}{\mbox{$\sigma^{\rm cloud}_{\rm mol}$}} % AL25 
\newcommand{\sigvcell}{\mbox{$\sigma_{\rm v}^{\rm cell}$}}
\newcommand{\sigvpm}{\mbox{$\mean{\sigma_{\rm mol}^{\rm cloud}}$}}

\newcommand{\sigvcellhex}{\mbox{$\mean{\sigma_{\rm v}^{\rm cell}}_{\rm hex}$}}
\newcommand{\rgal}{\mbox{$R_{\rm Gal}$}}
\newcommand{\rgalsun}{\mbox{$R_{\rm Gal,\odot}$}}
\newcommand{\reff}{\mbox{$r_{\rm e}$}}  % effective (half-light) radius
\newcommand{\msun}{\mbox{M$_\odot$}}% Msun
\newcommand{\tk}{\mbox{$T_K$}}

\newcommand{\n}{\mbox{$n$}}
\newcommand{\nbar}{\mbox{$\overline{n}$}}
\newcommand{\nbarcloud}{\mbox{$\nbar_{\rm cloud}$}}
\newcommand{\nbarcell}{\mbox{$\nbar_{\rm cell}$}}
\newcommand{\mmol}{\mbox{$M_{\rm mol}$}} % molecular mass

\newcommand{\mean}[1]{\mbox{$\langle#1\rangle$}} %generic mean for defined qu.
\newcommand{\av}{\mbox{$A_V$}} % Visual Extinction
\newcommand{\alphavir}{\mbox{$\alpha_{\rm vir}$}} % virial parameter
\newcommand{\avirp}{\mbox{$\alpha^{\rm cloud}_{\rm vir}$}} % virial parameter time PHANGS def
\newcommand{\avirpm}{\mbox{$\mean{\alpha^{\rm cloud}_{\rm vir}}$}} % virial parameter PHANGS def in hex
\newcommand{\avircell}{\mbox{$\alpha_{\rm vir, cell}$}} % virial parameter in cell
\newcommand{\avirhex}{\mbox{$\mean{\alpha_{\rm vir}}_{\rm hex}$}} % virial parameter in hex
\newcommand{\alphaco}{\mbox{$\alpha_{\rm CO}$}} % conversion L to M
\newcommand{\alphacoz}{\mbox{$\alpha_{\rm CO}(Z)$}} % conversion L to M fn of Z
\newcommand{\alphacosun}{\mbox{$\alpha_{\rm CO, \sun}$}} % conversion L to M for solar nbh
\newcommand{\alphacounit}{\mbox{\msun (K\ \kms\ pc$^2$)$^{-1}$}}
\newcommand{\xcounit}{\mbox{cm$^{-2}$ (K\ \kms)$^{-1}$}}
\newcommand{\epsff}{\mbox{$\epsilon_{\rm ff}$}} % efficiency per ff
\newcommand{\epsffhex}{\mbox{$\epsilon_{\rm ff}{\rm (hex)}$}} % efficiency per ff from tffhex/tdephex
\newcommand{\epsffEKO}{\mbox{$\mean{\epsilon_{\rm ff}{\rm (EKO)}}_{\rm hex}$}} % efficiency per ff from EKO and avg avir
\newcommand{\tff}{\mbox{$t_{\rm ff}$}} % free-fall time
\newcommand{\tffp}{\mbox{$\tau^{\rm cloud}_{\rm ff}$}} % free-fall time PHANGS def
\newcommand{\tffpm}{\mbox{$\mean{\tau^{\rm cloud}_{\rm ff}}$}} % free-fall time PHANGS def in hex
\newcommand{\tffcell}{\mbox{$t_{\rm ff, cell}$}} % free-fall time from mean density in cell
\newcommand{\tffhex}{\mbox{$\mean{t_{\rm ff}}_{\rm hex}$}} % free-fall time, M-weighted tffcell
\newcommand{\tdep}{\mbox{$t_{\rm dep}$}} % depletion time
\newcommand{\tdephex}{\mbox{$t_{\rm dep, hex}$}} % depletion time, Mhex/SFRhex
\newcommand{\ico}{\mbox{$I_{\rm CO}$}} % integrated intensity of CO
 
\newcommand{\hh}{\mbox{{\rm H}$_2$}}

\newcommand{\coo}{$^{13}$CO}
\newcommand{\cooo}{C$^{18}$O}

\newcommand{\jj}[2]{\mbox{$J = #1\rightarrow#2$}}
\newcommand{\go}{\mbox{$G_0$}}

\newcommand{\mstar}{\mbox{$M_{\star}$}}

\newcommand{\kkms}{\mbox{K\ \kms}}
\newcommand{\msunyr}{\mbox{M$_\odot$ yr$^{-1}$}}% Msun per year
\newcommand{\msunpc}{\mbox{M$_\odot$ pc$^{-2}$}}% Msun per square pc

\newcommand{\mw}{\mbox{\rm{Milky Way}}}

\shorttitle{The Milky Way as a galaxy I.}
\shortauthors{Evans et al.}
\graphicspath{{./}{../Analysis/}{../Analysis/Bigdips/}}

\defcitealias{2025ApJ...985...14L}{L25}
\defcitealias{MD17}{MD17}
\defcitealias{2016ApJ...822...52R}{R16}
\defcitealias{2025ApJ...980..216E}{E25} 

\begin{document}

\title{The Milky Way Joins the Extragalactic World: I. PHANGS}

\correspondingauthor{Neal J. Evans II}
\email{nje@astro.as.utexas.edu}

\author[0000-0001-5175-1777]{Neal J. Evans II}
\affiliation{Department of Astronomy, The University of Texas at Austin,
2515 Speedway, Stop C1400, Austin, Texas 78712-1205, USA}
\email{nje@astro.as.utexas.edu}

\author[0000-0002-9120-5890]{Davide Elia}
\affiliation{INAF-IAPS, Via del Fosso del Cavaliere 100, I-00133 Roma, Italy}
\affiliation{INAF, Sezione di Lecce, Via per Arnesano, I-73100 Lecce, Italy}
\email{davide.elia@inaf.it}

\author[0000-0003-0378-4667]{Jiayi Sun}
\affiliation{Department of Physics and Astronomy, University of Kentucky, 506 Library Drive, Lexington, KY 40506, USA}
\email{Jiayi.Sun@uky.edu}

\author[0000-0002-9333-387X]{Sophia Stuber}
\affiliation{National Astronomical Observatory of Japan, 2-21-1 Osawa, Mitaka, Tokyo 181-8588, Japan}
\affiliation{Max Planck Institute for Radio Astronomy, Auf dem Hügel 69, 53121 Bonn, Germany
}
\email{astro@sophiastuber.de}

\author[0000-0002-2545-1700]{Adam K. Leroy}
\affiliation{Department of Astronomy, The Ohio State University, 140 West
18th Ave, Columbus, OH 43210, USA}
\email{leroy.42@osu.edu}

\begin{abstract}
 Complete catalogs of molecular clouds in the Milky Way allow analysis of the molecular medium and the star formation properties of the Milky Way that closely follows the method used for nearby galaxies, in particular in the PHANGS project. The dependencies of the depletion time on other properties of molecular gas in the Milky Way are similar to those of other galaxies when analyzed in an analogous way. They exhibit a large scatter and relatively weak correlations. The strongest correlation is a decrease in depletion time with increasing velocity dispersion. We explore the effects of spatial resolution, sensitivity, cloud identificaton method, averaging method, and tracer choice on our results.
Metallicity effects in converting observations to mass can have a substantial impact. Somewhat fortuitously, differences in conversion methods between PHANGS and the Milky Way do not affect the current comparison. Inadequate spatial resolution, resulting in unresolved cloud structure, has the most important effect on interpretation of both extragalactic observations and existing catalogs of Milky Way clouds.

\end{abstract}

\keywords{ISM: molecules, ISM: clouds, star formation(1569)}

\section{Introduction} \label{sec:intro}

Our goal is to place the \mw\ in the context of recent comprehensive surveys of the molecular gas and star formation studies of large samples of nearby galaxies. Using similar analysis techniques provides insight into the limitations of the extragalactic data. This paper focuses on CO observations and the PHANGS-ALMA survey in particular. Another paper will focus on surveys of tracers of denser gas \citep{2016ApJ...822L..26B,2019ApJ...880..127J,2025A&A...693L..13N}. A spin-off of the current paper on the morphology of the \mw\ in comparison to barred galaxies was also published
\citep{2026ApJ...998L..23E}.

The PHANGS-ALMA survey
%Leroy et al. 2021
\citep{2021ApJS..257...43L}  
has provided data on the molecular gas and star formation rate (SFR) on $\sim 100$ pc scales for 90  galaxies.
The sample targeted nearby ($d < 20$ Mpc), low to moderately inclined ($i \lesssim 75\arcdeg$), relatively massive 
($\mstar/\msun \gtrsim \eten{9.75}$) star forming (${\rm SFR}/{\mstar} > \eten{-11}$ yr$^{-1}$) galaxies.

The PHANGS-ALMA team have explored many aspects of the data 
%S22
\citep{2021MNRAS.502.1218R,2022AJ....164...43S,2023A&A...678A.171Z,2022ApJ...927..149L,2024MNRAS.530..597B}.
The PHANGS-JWST project is providing infrared images of the sample
%Williams24 describes the survey and first release
\citep{2024ApJS..273...13W} and information on star formation, feedback, and dust physics \citep{2023ApJ...944L..17L}.
They have explored such issues as the duration of the embedded (not seen in visible light tracers) star formation
% Ramambason25
\citep{2026A&A...706A.186R}.
PHANGS-HST \citep{2022ApJS..258...10L} has obtained images of some galaxies with very high angular resolution, which  \citet{2025MNRAS.538.2744F} have used to create cloud catalogs based on extinction.
These studies provide a comprehensive view of molecular gas and star formation, but the  distance-limited spatial resolution may affect the conclusions. Comparison to studies in the \mw\ can provide valuable checks on that, so the goal is to add the \mw\ to the PHANGS data set. The \mw\ satisfies all the criteria for the PHANGS sample, except of course the inclination: $\mstar/
\msun = 10^{10.72}$ and ${\rm SFR}/\mstar = 3\ee{-11}$ yr$^{-1}$
% Imig, Licquia
\citep{2025ApJ...990..203I,2015ApJ...806...96L}.

Studies of other galaxies often make comparisons to the Milky Way, but they are hampered by the incomplete picture of our \mw\ caused by our location. Vast differences in spatial resolution across the \mw\ complicate analysis, and cloud properties are often biased toward nearby clouds. We now have quite complete maps of CO \jj10\ in the \mw\
% Dame 01
\citep{Dame01}
 and a cloud catalog that accounts for all the CO emission
% Miville-Deschenes 17
\citep{MD17}. 
%Elia22
\citet{2022ApJ...941..162E}
derived a face-on view of star formation in the \mw, using a model of star formation rate as a function of dense clump mass.
More recently,
%E25
\citet[][hereafter E25]{2025ApJ...980..216E}
have shown that three methods of predicting the total star formation rate and its distribution over Galactocentric radius (\rgal) provide consistent information. 

Additional surveys of the \mw\  are underway in different tracers with improved resolution, such as MWISP in CO, \coo, and \cooo\ \jj10\ 
\citep{2019ApJS..240....9S, 2020ApJS..246....7S,2025AJ....170..239Z}, 
SEDIGISM in \coo\ \jj21\
\citep{2021MNRAS.500.3027D, 2022MNRAS.513..296D, 2022A&A...664A..84N} and OGHReS in CO, \coo, and \cooo\ \jj21\ 
% Urquhart25 
\citep[][and references therein]{2025MNRAS.539.3105U}.
The time is right to place the \mw\ in the extragalactic context, at least in terms of its star formation and molecular gas properties.

Our primary comparison is to the paper by 
%Leroy et al. 2025
\citet[][hereafter L25]{2025ApJ...985...14L}, who 
have published a comprehensive study of cloud-scale gas properties, depletion times, and star formation efficiencies in a sub-sample of 67 galaxies from PHANGS-ALMA. They further provided a detailed, step-by-step description of their methods for computing the various properties of the clouds and star formation processes.
As much as is possible, we follow the same steps for the \mw. Some details differ because of the different nature of the datasets, but we specify these differences and consider the likely consequences. One difference is deliberate. Because our spatial resolution is much higher, we can examine the effects of spatial resolution on the outcomes, providing insights into the sub-grid assumptions used in the analysis of extragalactic data. To simulate the effects of the resolution of the PHANGS data, we assign each cloud in the \mw\ to a cell in a rectangular grid with a size equal to the PHANGS resolution.

%----------Table symbols------------------------
\begin{deluxetable*}{l c c c c c c l}
    \tablenum{1}
%    \tabletypesize{\footnotesize}
    \tablecaption{Symbol Definitions \label{tab:symbols}}
    \tablewidth{0pt}
    \tablehead{ 
    \colhead{Quantity} &  \colhead{PHANGS}   & \colhead{MW}   & \colhead{PHANGS}   & \colhead{MW}           & \colhead{PHANGS }  &  \colhead{MW}    &  \colhead{Note} 
}
%\decimalcolnumbers
    \startdata
        & \multicolumn{2}{c}{Clouds} & \multicolumn{2}{c}{Cells} & \multicolumn{2}{c}{Hexagons}  & \\
        \hline
Size & \nodata  & 0.1-70 pc & 150 pc & 150 pc & 1.5 kpc & 1.5 kpc & \\
Mass & \nodata  & \masscloud & $M_{\rm cl}$ & \masscell & $M_{\rm hex}$ & $M_{\rm hex}$ & \\
Free-fall Time & \nodata & \tff & \tffp  & \tffcell & \tffpm & \tffhex & 1 \\
Surface Density  & \nodata\ & \sigmamcloud\ & \sigmolcloud\ & \sigmamcell\ & \mean{\sigmolcloud} & \sigsmhex & 2 \\
Surface Density  & \nodata\ & \sigmamcloud\ & \sigmolcloud\ & \sigmamcloudmean\ & \mean{\sigmolcloud} & \sigmacloudhex & 3 \\
Velocity Dispersion ($\sigma_{\rm v}$) & \nodata\ & \sigvcloud\ & \sigvp\ & \sigvcell\ & \sigvpm & \sigvcellhex & 4 \\
Virial Parameter & \nodata & \alphavir & \avirp  & \avircell & \avirpm & \avirhex &  5 \\
Average Density (\nbar) & \nodata &  \nbar & \nodata & $\nbar_{\rm cell}$ & \nodata\ & $\mean{\nbar_{\rm cell}}_{\rm hex}$ & 6 \\
        \hline
    \enddata
%    \tablenotemark{$l$= Galactic Longitude, $b$= Galactic Latitude, \rg=Galactocentric Distance, }
% \tablenotetext{\dagger}{\cotw, \coo, \hcn, \hcop, BGPS} \tablenotetext{*}{\cotw, \coo, \hcn, \hcop, BGPS, \textit{Herschel} } }
\tablecomments{Notes apply to the column immediately to the left. 1. Averages are the inverses of mass-weighted averages over speeds, as defined by equation 1.
2. Mass-weighted mean of $\sigmamcell = M_{\rm cell}/A_{\rm cell}$.
3. Mass-weighted mean of \sigmamcloud\ for all clouds within the cell.
4. Mass-weighted mean of \sigvcloud\ for all clouds within the cell.
5. Mass-weighted mean of \avircell\ based on total size, mass, and velocity dispersion in a cell.
6. Mass-weighted mean of \nbarcell\ based on the mass and volume of a cell.
}
\end{deluxetable*}

Because many quantities are used, and the notation has to distinguish different kinds of averages, we provide for reference a table of symbols (Table \ref{tab:symbols}). For PHANGS, we generally follow their notation, but remind the reader that ``cloud" means 150 pc cloud-scale. In contrast, our cloud scale is much smaller and varies across the \mw. We use the term ``cell" for the 150 pc scale. Most of the quantities need no definition, but we clarify that \nbar\ is the mean density of all colliding partners; in a molecular cloud, this is $n(\hh) + n(\rm He) + \ldots $, calculated from
\begin{equation}\label{eq:nbar} 
\nbar = M/(V\,\mu\, m_{\rm amu}),
\end{equation}
where $M$ is the mass, $V$ is the volume, $m_{\rm amu}$ is the atomic mass unit. A standard assumption for the mean molecular weight, $\mu = 2.37$ \citep{2008A&A...487..993K}.
%Shirley et al. 2026]
\citet{2026PASP..138d3001S} has recently improved this value to $\mu = 2.3514\pm 0.0006$, for solar abundances, which we adopt. Variations with metallicity will be less than 1\%.

\section{Datasets}\label{sec:data}

\subsection{Extragalactic Data}\label{sec:exgaldata}

PHANGS-ALMA has produced a number of products that we use. Machine-readable tables recording high-level measurements from the various PHANGS datasets, described in
%Sun 22
\citet{2022AJ....164...43S},
were downloaded from CANFAR\footnote{
The data repository is 
\url{https://www.canfar.net/storage/list/AstroDataCitationDOI/CISTI.CANFAR/22.0072/data}
and we used NGC4321\_hexagon\_1p5kpc.ecsv from the v4p0\_public\_release for example.
}.
 Most of our comparisons are to plots and fits in
%Leroy 25 
%L25, 
\citetalias{2025ApJ...985...14L}
using tables published online with the paper. 
However, we use the tables published in the Erratum \citep{2026ApJ..1003..248L}, which correct some issues with the original tables.

In addition, the extinction-based cloud catalogs of
% Faustino Vieira et al.
\citet{2025MNRAS.538.2744F}
 provide a complementary view of structures, which is agnostic as to the composition, as both atomic and molecular gas is counted.
In particular, we use their catalog for NGC4321 to gauge the effect of using different tracers. It was one of the molecular morphological analogs for the \mw\ identified by \citet{2026ApJ...998L..23E}. While NGC4548 was even more similar to the \mw, it was not included in 
% Faustino Vieira et al.
\citet{2025MNRAS.538.2744F}.
For NGC4321, the spatial resolution allows detection of much smaller (radii defined by $\sqrt{A/\pi}$ down to 8 pc, where $A$ is the area) and less massive (minimum mass of 1.5\ee3 \msun) clouds than those identified from the ALMA data. From the directory,
/AstroDataCitationDOI/CISTI.CANFAR/25.0036/data,
we downloaded the catalog for NGC4321 (NGC4321\_cloud\_catalogue.ecsv) and the homogenized catalog of all four galaxies (homogenised\_cloud\_catalogue.ecsv).

\subsection{Milky Way data}\label{sec:mwdata}

The data on molecular gas in the \mw, for comparison to PHANGS, comes from the catalog of clouds defined by CO \jj10\ emission
\citep[][hereafter MD17]{MD17}.
%E25
%\citet[][hereafter E25]{2025ApJ...980..216E} 
%Elia et al. 2025
\citetalias{2025ApJ...980..216E}
adjusted this catalog to use the same rotation curve as was used by the Hi-GAL survey, adjusted masses for a value of \alphaco\ that varied with metallicity, $Z$, and provided star formation rates predicted by the model of 
\citet[][hereafter EKO]{2022ApJ...929L..18E},
using a star formation efficiency per free-fall time (\epsff) that varied with the virial parameter (\alphavir), as found by \citet{2021ApJ...911..128K}.

For a few comparisons, we also use the catalog of
%Rice
\citet[][hereafter R16]{2016ApJ...822...52R}, which is less complete (includes about 25\% of the total CO \jj10\ emission of the \mw) and discriminates against low mass clouds compared to the catalog of 
%MD17
\citetalias{MD17}. It is useful to study the effects of different cloud identification methods.

We use the catalogs of sources rather than going back to the input data. This avoids repeating past work. An exception is the measure of \sfr. Because the \sfr s in the catalog of 
%E25
\citet{2025ApJ...992..161E}
are based on the model of EKO, which assume a relation between \sfr\ and cloud mass, free-fall time, and virial parameter, tests of the dependence of \tdep\ on cloud properties would be circular. 
%Elia 25
However, 
%E25
\citetalias[]{2025ApJ...980..216E}
showed that the \sfr\ based on an alternative approach, consisting in converting the \textit{Herschel} emission at 70~\micron\ into a luminosity  at that wavelength (by using the heliocentric distance of the nearest compact source), and hence into \sfr\ according to the prescription of \citet{2010ApJ...725..677L}, gave a total \sfr\ and dependence on \rgal\ that is very similar to that of \sfr\ based on EKO. Consequently, we use the SFR based on the whole 70 \micron\ emission, after converting pixel intensities into luminosities through distances assigned to the closest  clumps, as described in %E25
\citetalias{2025ApJ...980..216E}.
The resulting surface densities of \sfr\ and molecular gas satisfy a Kennicutt-Schmidt relation with a power-law exponent of $1.10 \pm 0.06$, identical within uncertainty to that found for the PHANGS galaxies \citep{2023ApJ...945L..19S}.

\section{Procedures}\label{sec.proc}

Here we outline the procedures used by 
%Leroy et al. 2025
%L25
\citetalias{2025ApJ...985...14L}
and how we depart from them. ``PA" indicates the PHANGS-ALMA step and ``CP" indicates the current paper. The issue of the luminosity to mass conversion factor (\alphaco) and its dependence on metallicity are discussed in \S \ref{sec:alphaco}.

\begin{enumerate}
\item (PA) Choose a threshold of CO \jj21\ emission and apply a local, best-estimate mass conversion factor (\alphaco) to estimate the mass on the scale of 150 pc. The threshold value of integrated intensity is 0.6 \kkms\ to achieve a uniform sensitivity limit.

(CP) We have CO \jj10\ emission, rather than \jj21\ data. We have cloud catalogs rather than images. The threshold for the \mw\ data is lower.
For our standard procedure, we apply no criterion on integrated intensity, but we adopt the requirements for clouds used by 
%Elia25
%E25
\citetalias{2025ApJ...980..216E}
as follows:
$\rgal \leq 20$ kpc, $\mmol \geq 1$ \msun, $\alphavir < 100$. Restrictions on CO integrated intensity are then added to test the effects of sensitivity limitations (\S \ref{sec:sensitivity}).
The ``best-estimate" value of \alphaco\ will differ because of the different transition, but this difference turns out to be small (\S \ref{sec:alphaco}).

\item (PA) Convolve the CO to a resolution of 150 pc (FWHM). Estimate the surface density and effective line width, correct for inclination to calculate face-on values, and calculate virial parameter, \alphavir, and free-fall time, \tff. Following
%Sun 2020
\citet{2022AJ....164...43S},
they use the following definition of average free-fall time:
\begin{equation}\label{eq:tffL25}
\mean{\tff} = \bigg(\frac{\sum_i M_i (\tff_i)^{-1}}{\sum_i M_i}\bigg)^{-1} ,
\end{equation} 
where $M_i$ and $\tff_i$ are the mass and free-fall time of object $i$. This method actually averages speeds of formation and weights by mass, then converts to \tff. 
The value of $\tff_i$ is calculated from
\begin{equation}\label{eq:tff}
\tff_i = \sqrt{3 \pi/(32 G \rho)},
\end{equation}
and 
\begin{equation}\label{eq:rho}
\rho = 3 M/(4 \pi R^3_{\rm pix}).
\end{equation}
Finally, 
\begin{equation}\label{eq:Rpix}
R_{\rm pix} = {\rm min}[l/2, (l^2 H/(8 \cos i))^{1/3}],
\end{equation}
where $l$ is the side of the cell, $H$ is the scale height of molecular gas (taken to be 100 pc) and $i$ is the inclination angle of the galaxy. 

(CP) We have catalogs of these properties already and higher resolution.
 Each cloud is assigned to a rectangular cell, 150 pc on a side, based on the Galactic coordinates of its centroid and its distance. Clouds are not split between cells. The cloud distance is the primary source of uncertainty, especially in regions where kinematic distances are questionable, such as the region of the bar \citep{2026PASJ...78.1093B}. The procedure is described in more detail in \citetalias{2025ApJ...980..216E}, where it was used for 500 pc cells.
 
While \citetalias{2025ApJ...985...14L} defines $\tff_i$ as the free-fall time of ``cloud" $i$, that definition is for 150 pc resolution; for comparison, we use \tffcell, calculated from the mean density of a cell (mass of cell divided by volume). For the \mw,  $\cos i = 1$, and $R_{\rm pix} = 65.5$ pc. We calculate both the average of the cloud level properties from our catalog and the smoothed value based on resolution of the 150 pc cell, labeled ``cell." For example, the average surface density of all the clouds in a cell is $\mean{\Sigma_{\rm cloud}}$, while the smoothed value is $\Sigma_{\rm cell} = M_{\rm cell}/A_{\rm cell}$, where $M_{\rm cell}$ is the total mass of all the clouds in the cell and $A_{\rm cell}$ is the area of the cell (2.25\ee4 pc$^{2}$). The latter is the equivalent of $\Sigma^{\rm cloud}_{\rm mol}$ of \citetalias{2025ApJ...985...14L} while the former is not available in PHANGS because of resolution. Note that the superscript ``cloud" in the \citetalias{2025ApJ...985...14L} notation denotes  cloud-scale rather than cloud, but we will use cell because many clouds exist in a typical 150 pc cell. One issue with using the catalog, rather than the original CO data arises in the linewidth. We use only the linewidth of clouds, averaged (unweighted) over the cell, which does not take into account velocity differences of clouds within the cell, so we are less likely to overestimate the velocity dispersion within a cell. The virial parameter of the cell is computed from the total molecular mass $M_{\rm cell}$ and radius $r_{\rm cell} = 65.5$~pc to match the L25 method, along with the velocity dispersion \sigvcell\ (so it could be underestimated as well):

\begin{equation}\label{eq:avircell}
\avircell = 5 r_{\rm cell} \sigvcell^2/(G M_{\rm cell}) .
\end{equation}

\item (PA) Divide the galaxy into hexagonal apertures with diameter 1.5 kpc. Within each aperture, compute the mass-weighted expectation value of the surface density, velocity dispersion and virial parameter from the ``cloud" level values (really the 150 pc cell values).

(CP) We also divide the \mw\ into 1.5 kpc hexagons and compute average values of properties within those hexagons. Our standard procedure also uses mass-weighting, where mass is the total mass of the cell, $M_{\rm cell}$, to best replicate the procedure in \citetalias{2025ApJ...985...14L}. The results are labeled with ``hex," as in $\mean{\Sigma_{\rm cell}}_{\rm hex}$ for the mass surface density. Because we have information on individual clouds, we also compute the average of cloud surface densities, again mass weighted by the mass in the cell: $\mean{\Sigma_{\rm cloud}}_{\rm hex}$.
For consistency with our limitation of $\rgal < 20$ kpc, we restrict the centers of the hexagons. At this stage, we add information on the 70 \micron\ \sfr, averaged over the hexagon.
Figure \ref{fig:hexagons} shows the hexagons with the surface densities of molecular gas and star formation rate from 
%E25
\citetalias{2025ApJ...980..216E} in color code. For further analysis, each hexagon must have non-zero molecular mass. This requirement cuts the number of hexagons to 316. The total \sfr\ of the \mw\ decreased only slightly from 1.41 to 1.21 \msunyr, indicating that, according to distance assignments done in 
%E25
\citetalias{2025ApJ...980..216E}, little 70~\micron\ emission resides outside hexagons with molecular clouds. Comparison of the two panels of Figure \ref{fig:hexagons} shows that most of the hexagons without molecular mass are on the far side of the \mw.

\begin{figure*}[ht!]
\includegraphics[width= 0.50\textwidth]{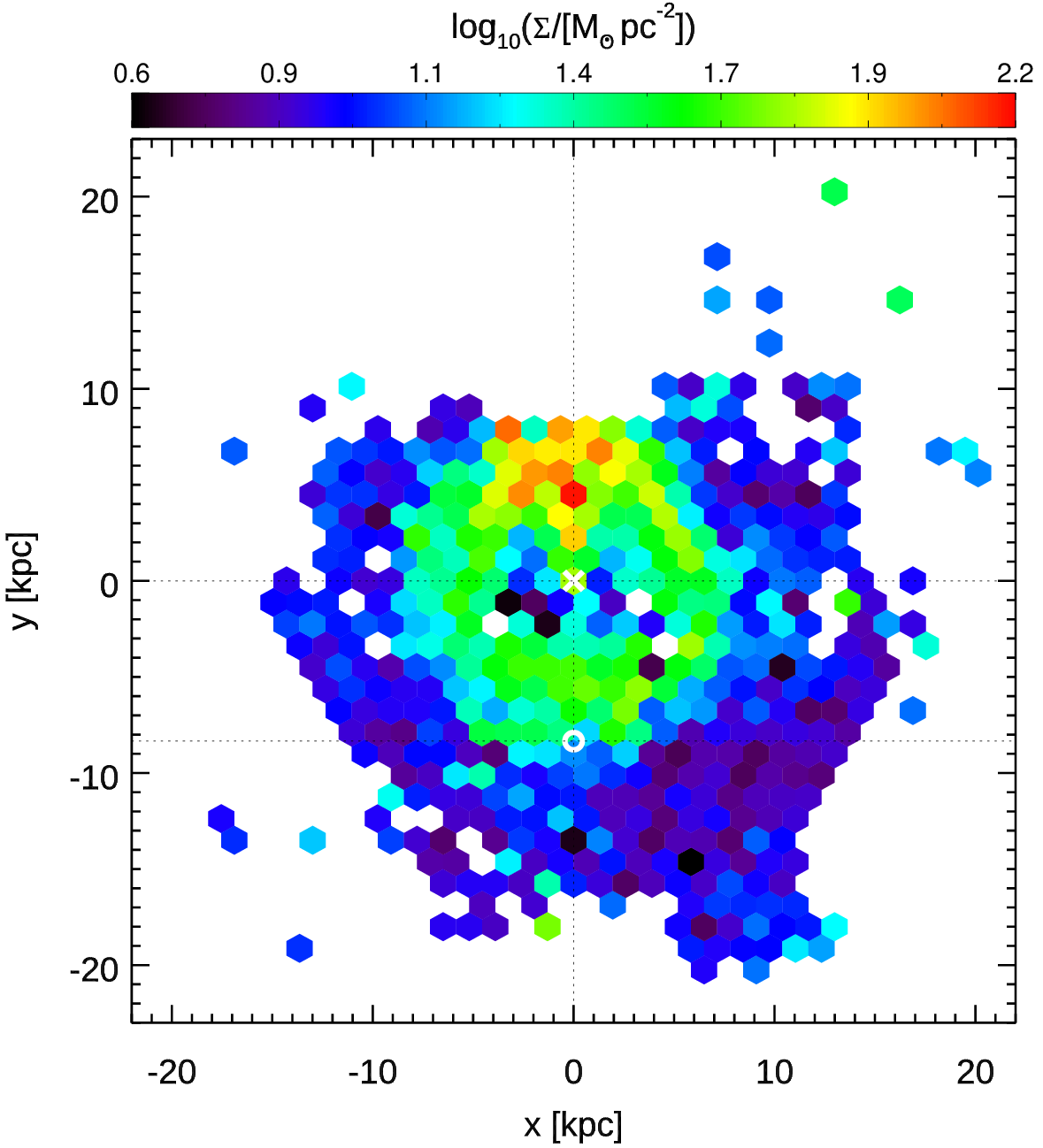}
\includegraphics[width= 0.50\textwidth]{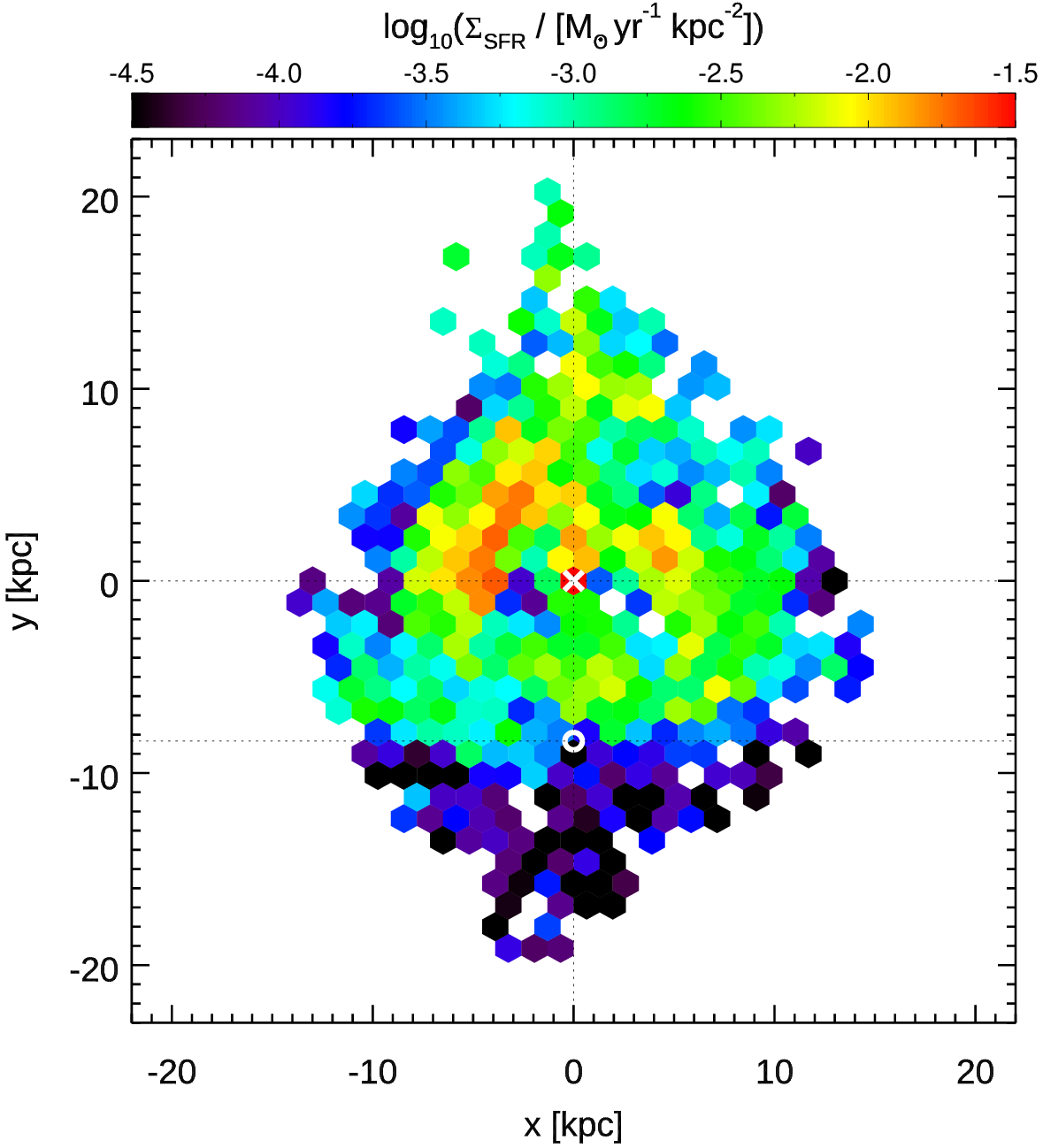}
\caption{Layout of the hexagonal bins showing the molecular gas surface density  (left) and the 70 $\mu$m SFR surface density (right). The right panel presents the same data as Figure 5 (left) of 
%E25
\citetalias{2025ApJ...980..216E}, but re-binned from the original rectangular grid into hexagons. The white circle and X mark the positions of the Sun and the Galactic center, respectively, further highlighted by the intersections of the displayed vertical and horizontal dotted lines. 
}
\label{fig:hexagons}
\end{figure*}

\item (PA) Calculate the molecular gas depletion time within the 1.5 kpc hexagons by convolving $\Sigma_{\rm SFR}$ and $\Sigma_{\rm mol}$ to 1.5 kpc resolution and dividing them.

(CP)  We used the prediction of star formation rate from the 70 \micron\ values for each hexagon. Then, we computed the depletion time in each hexagon, \tdephex, from the  total star formation and mass in each hexagon. This seems the closest approach to recreating the L25 procedure, although we do not exactly match the convolution process.

\item (PA) Divide the average cloud \tff\ by \tdep\ to estimate the efficiency per free-fall time for each 1.5 kpc region. 

(CP) We do the same with the definition of \tff\ in equation 2. We explore the effects of using different definitions in Appendix \ref{sec:tffdef}. Using our notation, $\epsffhex = \tffhex/\tdephex$.

\item (PA) Limit the analysis to regions with surface densities at the hex-scale ($\mean{\sigmolcloud}$, see Table \ref{tab:symbols}) greater than 20 \msunpc. 

(CP) We apply the same cut to our hexagons, using $\mean{\Sigma_{\rm cell}}_{\rm hex}$, which is our equivalent.  At this stage, we also require a non-zero value for the 70~\micron\ \sfr\ in the hexagon. This step cuts the number of hexagons  to 195, eliminating many of the outer \mw\ hexagons (Figure \ref{fig:hexagons}). The total 70~\micron\ \sfr\ drops  only to 1.10 \msunyr, indicating that there is little star formation at lower surface densities.

\end{enumerate}

We create three products for each property: the average of the cloud level properties from the catalog, grouped into 150 pc cells, labeled ``cloud"; the value based on resolution of the 150 pc cell, labeled ``cell"; and the mass-weighted average over each 1.5 kpc hexagon, labeled ``hex." The last two are most directly comparable to the PHANGS-ALMA results, while the first tests the effects of the limited resolution and sensitivity. See Table \ref{tab:symbols} for definitions of symbols used by \citetalias{2025ApJ...985...14L} and here.

As a result of these procedures, we have values for the \mw\ that have been derived in a way that mimics the procedures used for other galaxies. We will also have information on the effect of spatial resolution on the extragalactic analysis.

Appendix \ref{app:details} addresses relatively simple issues of how to compare the two datasets. In \S \ref{sec:prelims}, we discuss more complex or consequential issues.

\section{Preliminary Considerations}\label{sec:prelims}

\subsection{Luminosity-to-Mass Conversion}\label{sec:alphaco}

The mass of molecular gas is related to the luminosity of a CO line by the conversion factor, \alphaco\ with units \alphacounit. Until recently, studies of the \mw\ have used a constant value for \alphaco, but recent studies covering a wider range of \rgal\ have corrected for metallicity effects
% EKO, E25
\citep{2022ApJ...929L..18E,2025ApJ...980..216E}.
In this regard, the extragalactic community has been ahead, faced with galaxies of widely different metallicities 
% Bolatto 13
\citep{2013ARA&A..51..207B}
and excitation conditions 
%Downes and Solomon 98
\citep{1998ApJ...507..615D}.
In the most recent authoritative formulation,
% Schinnerer and Leroy 24
\citet{2024ARA&A..62..369S}
favor the following formula:
\begin{equation}\label{eq:aco}
\alphaco = \alphacosun f(Z) g(\sigmastar),
\end{equation}
where $Z$ is the metallicity, $f(Z)$ corrects for metallicity effects, \sigmastar\ is the surface density of stars, and $g(\sigmastar)$ corrects for the enhancement of CO emission in regions of high stellar density, like centers of galaxies.\footnote{They use $\alpha^{1-0}_{\rm CO,MW}$, but we replace ``MW" with the Sun symbol to mean the value determined in the solar neighborhood because there is no single value for the \mw, and we drop the superscript because our data are all for \jj10.} We refer to this model as SL.

The factor $f(Z)$, which they refer to as the CO-dark factor, has been much debated. 
%Schinnerer and Leroy
\citet{2024ARA&A..62..369S}
recommend 
\begin{equation}\label{eq:f(Z)}
f(Z) = (Z/Z_\sun )^{-a}
\end{equation}
and favor $a = 1.5$ in the range of 0.2 to 2.0 solar.
We define $Z$ to be relative to solar, so here we use $f(Z) = Z^{-a}$.

The factor $g(\sigmastar)$, called the starburst factor, accounts for the fact that clouds in galactic centers are more luminous per unit mass
% Schinnerer and Leroy 24
\citep{2024ARA&A..62..369S}. 
\begin{equation}\label{eq:g}
g(\sigmastar) = \bigg(\frac{\max(\sigmastar,100\ \msunpc)}{100\ \msunpc}\bigg)^{-0.25} \,.
\end{equation}
It is only relevant for $\sigmastar > 100$ \msunpc. We estimate its effects in Appendix \ref{sec:starsurfden}. It only affects the inner 2 kpc of the \mw\ and by less than a factor of 2.

In addition, recent papers including \citetalias{2025ApJ...985...14L} use a variable $R_{21}$ to convert from \jj21\ to \jj10\ intensity that depends on the surface density of star formation
%S25
\citep{2025ApJ...994..263S}: 
\begin{equation}\label{eq:R21}
R_{21} = 0.65 (\sigmasfr/0.018)^{0.125},
\end{equation}
with \sigmasfr\ measured in \msun\ yr$^{-1}$ kpc$^{-2}$.
This is not relevant to the \mw, for which the surveys are in the \jj10\ transition.

While many references still use a constant \alphaco\ for the \mw, we have adopted for our recent papers 
%(E25) 
\citepalias{2025ApJ...980..216E}
a metallicity-dependent conversion factor, 
\begin{equation}\label{eq:acoE25}
\alphacoz = 4.50 Z^{-0.8}
\end{equation}
with the exponent based on simulations 
% Gong, Hu
\citep{2020ApJ...903..142G, 2022ApJ...931...28H}. 
Previous studies in the MW show that $a = 0.8$ works best when using the \mw\ gradient in oxygen abundance 
% EKO , Elia25
\citep{2022ApJ...929L..18E,2025ApJ...980..216E}.
The common choice for \alphacosun\ is 4.35 \alphacounit, based on adding helium, when converting the local ``standard" $X_{\rm CO}$ of 2\ee{20} \xcounit. If all the elements are added, the correct value is $\alphacosun = 4.50$
% EKO,
\citep{2022ApJ...929L..18E},
which we have adopted for the \mw. 

\begin{figure}[ht!]
\includegraphics[width=0.45\textwidth]{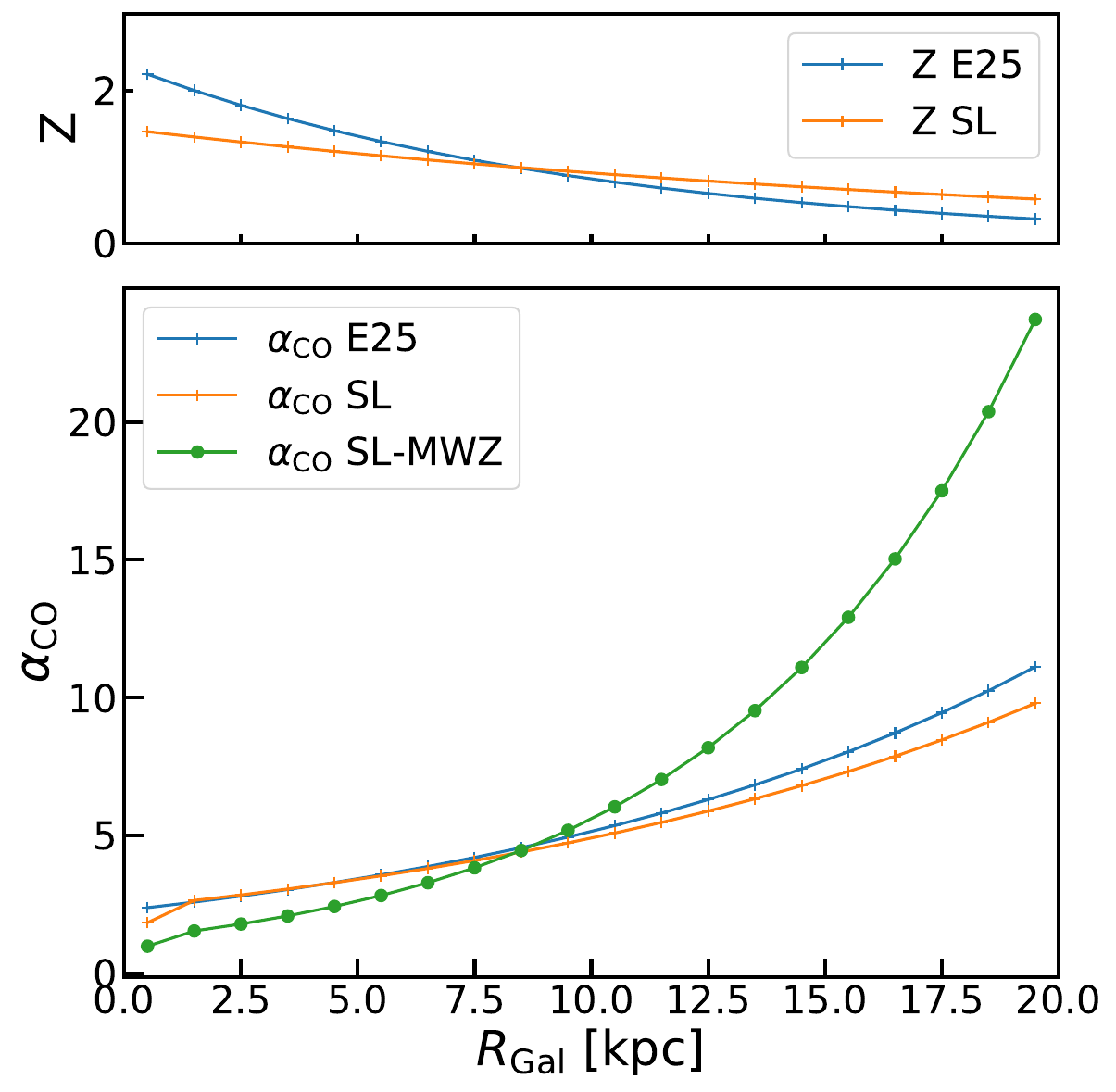}
\caption{ Two models for \alphacoz\ are plotted versus \rgal\ (bottom), along with two models for $Z(\rgal)$ (top). The \mw\ model for $Z(\rgal)$ assumes a gradient of $-0.044$ dex/kpc extending throughout the \mw\ (labeled E25). The model labeled SL uses the extragalactic model with a gradient of $-0.10/r_{\rm e}$ dex. The model labeled SL-MWZ uses those models for \alphacoz\ but the measured [O/H] gradient in the \mw\ 
% Mendez-Delgado 22
\citep{2022MNRAS.510.4436M}.}
\label{fig:alphaz}
\end{figure}

While the SL formulae for \alphacoz\ depend much more strongly on $Z$ than those we have used for the \mw, the effect is muted by the weaker dependence of $Z$ on \rgal\ used in PHANGS papers.
 For the \mw\ we used a measured oxygen abundance gradient of $-0.044 \pm 0.009$ dex/kpc \citep{2022MNRAS.510.4436M}. Other elements show even larger gradients. Lacking a consistent set of metallicity measurements for the full sample, the PHANGS team adopted metallicities from scaling relations, as described in detail by \citet{2022AJ....164...43S}. In particular, their gradient is normalized to the effective (half-light) radius of the galaxy (\reff):
\begin{equation}\label{eq:Z(Rgal)}
\log Z(\rgal) = \log Z(\reff)  - 0.1 \rgal/\reff
\end{equation}
with $\log Z(\reff)$ a function of total stellar mass
%Sanchez 14, 19
\citep{2014A&A...563A..49S,2019MNRAS.484.3042S}.
\begin{equation}\label{eq:Z(Reff)}
\begin{split}
\log Z(\reff) ={}& 0.04 + 0.01(\log(\mstar/\msun) - 11.5) \\  
&\times \exp(-\log(\mstar/\msun) + 11.5) .
\end{split}
\end{equation}
We refer to the combination of \alphacoz\ and $Z(\rgal)$ models as SL in Figure \ref{fig:alphaz}.  

To apply the PHANGS formulae to the \mw, we need to know \mstar\ and \reff.
Our knowledge of the stellar properties of the \mw\ is in a state of rapid change. Values for the total stellar mass differing by a factor of two have been published recently 
\citep{2025ApJ...990..203I,2025ApJ...990L..37L}, along with estimates of \reff\ ranging from 4.12 kpc
% Lian 24
\citep{2025ApJ...990L..37L}
to 4.75 kpc
% Imig 25
\citep{2025ApJ...990..203I} 
to 5.75 kpc
% Lian24
\citep{2024NatAs...8.1302L}
We adopt $\reff = 4.75$ kpc and $\log \mstar = 10.72$ 
%Imig25
\citep{2025ApJ...990..203I}
 for this purpose, yielding a gradient of $-0.021$ dex/kpc, less than half the measured value, as can be seen in the top panel of Figure \ref{fig:alphaz}.\footnote{For discussion of \reff\ for the \mw, see Appendix \ref{sec:starsurfden}.}
For comparison to extragalactic gradients, the measured metallicity gradient in the \mw\ of $-0.044$ dex/kpc and $\reff = 4.75$ kpc implies $-0.209$ dex/\reff, larger than in any of the 19 PHANGS galaxies studied by
%Groves 23
\citet{2023MNRAS.520.4902G}.
Those are characterized by a mean of $-0.059 \pm 0.056$ dex/\reff, and the most negative value is $-0.188 \pm 0.005$ for NGC1365.
These formulae used for extragalactic work predict $Z(\reff) = 1.05$ and $Z(\rgalsun) = 0.70$, where we take $\rgalsun = 8.34$ kpc, for consistency with %E25
\citetalias{2025ApJ...980..216E}.
The SL formula would be in serious disagreement  with the metallicity gradient for the \mw, measured in nebular lines
% Mendez-Delgado 22
\citep{2022MNRAS.510.4436M}.
% ref from Keith
Studies of stars also support a steeper metallicity gradient in the \mw\
\citep{2017A&A...600A..70A}, than what would be inferred from the SL formulae.

The various formulations for \alphaco(\rgal) are plotted in the lower panel of Figure \ref{fig:alphaz}. The different formulations for $Z(\rgal)$ result in rather similar \alphaco(\rgal) at large \rgal\ if we use the SL models of $Z(\rgal)$; they would be very different if we used the SL formula for \alphacoz\ and the actual $Z(\rgal)$ measured for the \mw\ (green curve). The sharp-eyed reader may notice a downturn for the innermost point on the SL plots, caused by the $g(\sigmastar)$ factor.

When comparing \mw\ data to the PHANGS data, we will use the same conversion methods
%SL 24
\citep{2024ARA&A..62..369S}
 that were used in the PHANGS data for consistency.  As Figure \ref{fig:alphaz} shows, the difference between these conversion methods and the one we believe applies to the \mw\ is very small. However, this seems a rather fortuitous cancellation of two effects.

\subsection{Effects of Resolution}\label{sec:resolution}

We consider the effects of the limited spatial resolution for extragalactic observations by sorting the \mw\ data into cells of 150 pc on a side, simulating the resolution of the PHANGS-ALMA data. We compare various properties when we average over the resolved clouds in the catalog compared to aggregating them into a single mass.

%----------Table Resolution and Sensitivity----------   
\begin{deluxetable*}{l r r r r r r r r r r r r } 
\tablecaption{Effects of Resolution and Sensitivity \label{tab:res}} 
\tablewidth{0pt} 
\tablehead{   
\colhead{Sample } &  \colhead{Number}   & \multicolumn{3}{c}{$\log (\nbar/\cmv$)}   & \multicolumn{3}{c}{\sigmam\ (\msunpc)}   & \multicolumn{3}{c}{\alphavir }   & \colhead{$\log (M_{\rm tot}/\msun$ )} & \colhead{Note} \\  
 & & Med & Mean & Std. & Med & Mean & Std.  &     Med & Mean & Std.          &        &  
  } 
\startdata 
\hline 
MW-E25-cloud & 7389 & 1.05 & 1.07 & 0.65 & 17.9 & 28.7 & 29.4 & 8.03 & 15.41 & 18.40 & 9.13 & 1 \cr 
MW-E25-cell & 4558 & 0.82 & 0.86 & 0.57 & 4.3 & 13.1 & 27.1 & 10.88 & 36.11 & 117.44 & 9.13 & 2 \\ 
MW-E25-hex & 402 & 0.63 & 0.70 & 0.51 & 13.2 & 27.6 & 40.8 & 7.22 & 11.98 & 19.63 & 9.12 & 3 \\ 
MW-L25-cloud & 6575 & 1.14 & 1.14 & 0.64 & 21.2 & 31.7 & 29.8 & 7.50 & 14.52 & 17.71 & 9.12 & 4 \cr 
MW-L25-cell & 2025 & 1.34 & 1.34 & 0.43 & 11.1 & 22.2 & 36.0 & 7.85 & 19.79 & 76.30 & 9.00 & 5 \\ 
MW-L25-hex & 195 & 1.20 & 1.17 & 0.40 & 33.0 & 45.7 & 51.0 & 5.47 & 7.91 & 11.63 & 9.00 & 6 \\ 
\enddata  
\tablecomments{ 1. Data from catalog in  \citet{2025ApJ...980..216E} using their criteria: $\mmol > 1$ \msun, $R_{\rm gal} < 20$ kpc,  and $\alphavir < 100.$ \\ 2.  Same as 1 but assuming resolution of a 150 pc cell. \\ 
  3. Same as 1 but based on mass-weighted averages in 1.5 kpc hexagons.  \\ 
 4. Data from catalog in \citet{2025ApJ...980..216E} adding criterion on $\ico > 1.0$ \kkms. \\ 
 5. Same as 4 but assuming resolution of a 150 pc cell.  \\  
 6. Same as 4 but based on mass-weighted averages in 1.5 kpc hexagons. \\ 
 } 
 \end{deluxetable*}

\begin{figure*}[ht!]
\includegraphics[width=0.50\textwidth]{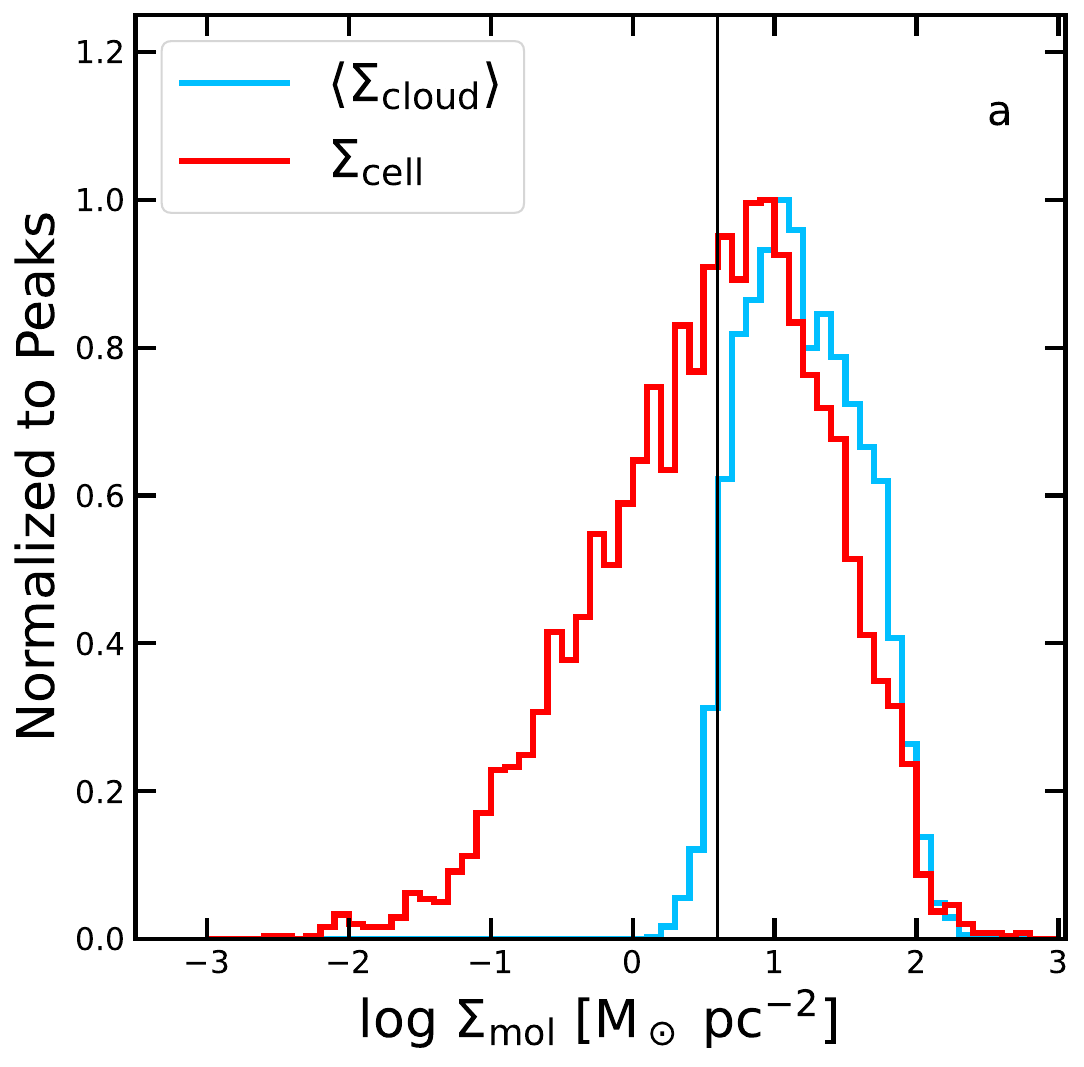}
\includegraphics[width=0.50\textwidth]{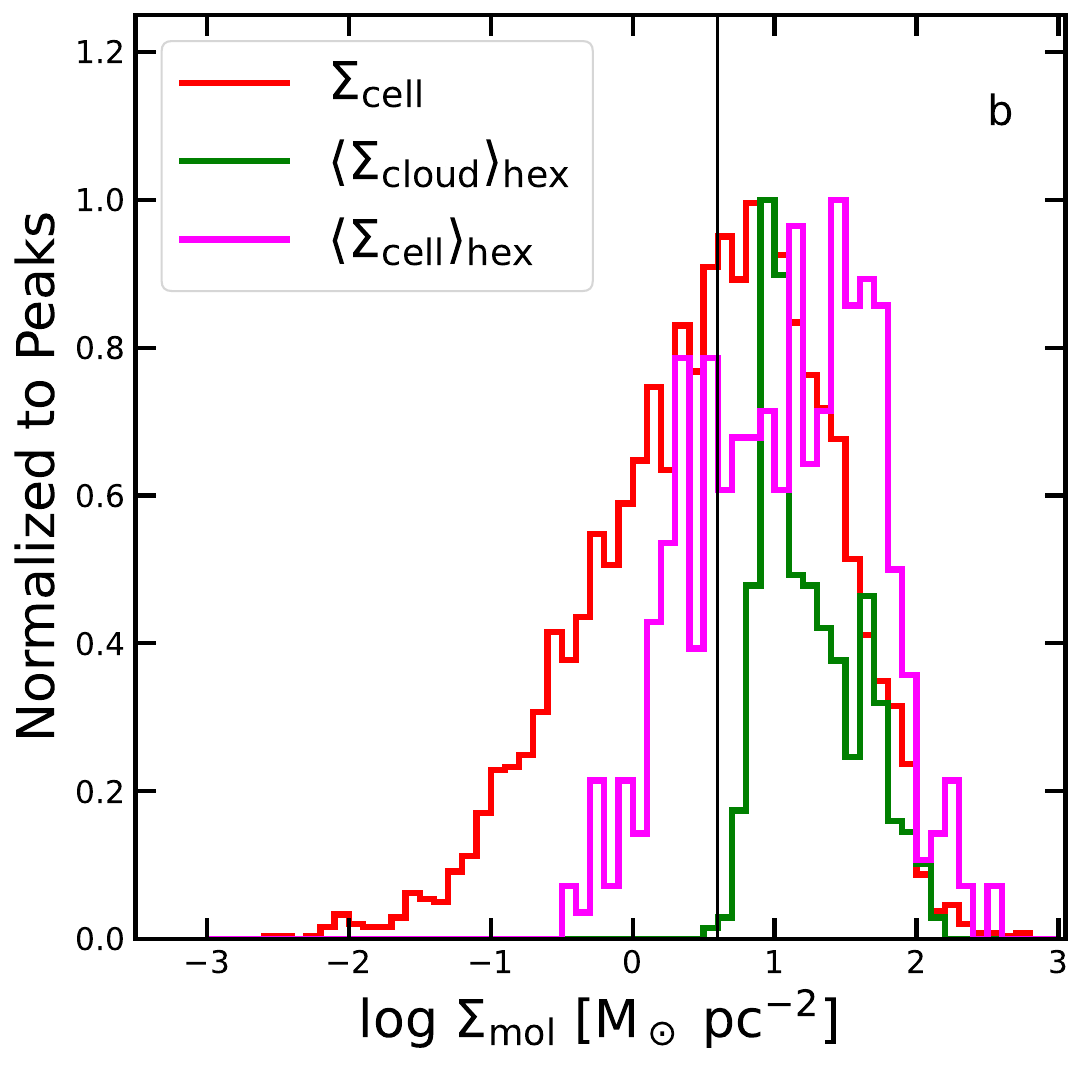}
\caption{(a) The logarithmic mass surface density distribution for CO clouds calculated two ways, both normalized to the peak value.  The cyan histogram is the distribution of \sigmamcloud\  of clouds averaged over the cell (\sigmamcloudmean\ in Table 1). The red histogram is the ratio of the total mass in the cell to the area of the cell (\sigmamcell\ in Table 1). 
(b) The distribution of the logarithm of the (mass) surface density distribution for CO clouds calculated three ways, all normalized to the peak value.  The red histogram is the ratio of the total mass of the cell to the area of the cell ($\Sigma_{\rm cell}$), repeated from panel a. The green histogram is the mass-weighted average of the cloud level surface density ($\Sigma_{\rm mc}$) over the hexagons (\sigmacloudhex\ in Table 1). The magenta histogram is the mass-weighted average of $\Sigma_{\rm cell}$ over the hexagons (\sigsmhex\ in Table 1). No sensitivity limit is used for the sample. The vertical line is $\log (4~\msunpc)$, the sensitivity limit of PHANGS-ALMA.}
\label{fig:res1}
\end{figure*}

The surface density (\sigmam) is often used to characterize molecular clouds, the volume density directly predicts the free-fall time, and the virial parameter summarizes the dynamical state. We compare in  Table \ref{tab:res} the distribution of these three quantities for the 
%MD17
\citetalias{MD17} sample, sorted into cells if we average the \sigmam\ of the clouds (denoted \sigmamcloud) versus dividing the total mass in the cell by the area of the cell (denoted \sigmamcell), roughly approximating the extragalactic resolution. The low resolution clearly lowers the values of \sigmam\ by about a factor of three (cf. entries 1 and 2 in Table \ref{tab:res}).
 The tail to very low \sigmamcell\ in Figure \ref{fig:res1} reflects cells with very few or only low mass clouds; these would escape detection by PHANGS-ALMA altogether, since their limit is 4 \msunpc, as shown by the vertical line.  In contrast, mass-weighted averaging over hexagons increases the surface density as shown by line 3 in Table \ref{tab:res}, where \sigsmhex\ is very similar to the actual average cloud surface density (\mean{\sigmamcloud}), and the distribution of \sigsmhex\ is closer to the actual distribution of \sigmamcloud\ (Fig. \ref{fig:res1}). An extragalactic observer using 150 pc resolution to observe the \mw\ would underestimate the mean cloud-level surface density in a 150 pc beam, but slightly overestimate the mean surface density by mass-weighting the data into hexagons.

%\begin{figure}[ht!]
%\includegraphics[width=0.45\textwidth]{E25res2.pdf}
%\caption{The distribution of the logarithm of the (mass) surface density distribution for CO clouds calculated three ways, all normalized to the peak value. No sensitivity limit is used for the sample.   The vertical line is $\log (4~\msunpc)$, the sensitivity limit of PHANGS-ALMA.}
%\label{fig:res2}
%\end{figure}

\subsection{Effects of Sensitivity}\label{sec:sensitivity}

We next ask how the \mw\ properties change if we institute cuts that simulate the sensitivity limits of PHANGS-ALMA. We compare basic results with the full catalog and with cuts at the cloud level where we require an intensity of CO \jj10\ emission of 1.0 \kkms, equivalent to their \jj21\ sensitivity of 0.65 \kkms\ (see Appendix \ref{app:details} for details).
Only 814  clouds fail the CO intensity criterion, with a total mass of 6\ee7 \msun. Table \ref{tab:res} in lines labeled ``MW-L25" provides comparisons of the median, mean, and standard deviation of the average densities (\nbar), surface densities (\sigmam), and virial parameters (\alphavir). Adding the criterion on \ico\ corresponding to the PHANGS-ALMA sensitivity limits raises the values of \nbar\ and \sigmam\  and lowers the values for \alphavir\ for the \mw\ clouds because less populated cells are left out of the averages, but the effect is very small.

  We use only the criteria for clouds used by 
%E25
\citetalias{2025ApJ...980..216E}  for the next set of comparisons.

\subsection{Tracers and Cloud Identification Methods}\label{sec:tracers}

Because different tracers (CO~\jj10, CO~\jj21, extinction) and different cloud separation methods have been used to define clouds, we compare the properties from some characteristic references. These both affect the definition of a cloud.
For example,
%Wang et al 25
\citet{2025ApJS..280...16W} have produced a catalog of \mw\ clouds based on extinction, but favoring the denser, smaller leaves on a dendrogram analysis. The comparison to properties from CO and other dust-based catalogs emphasizes the inherent ambiguities in cloud definition.

We explore first the properties of the samples of clouds in the \mw, as compared to a similar PHANGS galaxy. We choose NGC4321. This is a barred galaxy, with data in both PHANGS, as well as \citet{2025MNRAS.538.2744F}, who derive cloud properties from extinction. For the properties measured from CO~\jj21, we restrict the entries in \citetalias{2025ApJ...985...14L} Table~2 \citep{2026ApJ..1003..248L} to this galaxy. Because density and mass are not given, we calculate the total mass and \nbar, from the surface density, assuming a volume of the hexagon with a total thickness of 100~pc.
For NGC4321 measured in extinction \citep{2025MNRAS.538.2744F}, we get mass from ``Mass", radius from ``R\_eq", and area from ``Area\_exact".
As explained in 
%Vieira 23
\citet[][equation 7]{2023MNRAS.524..161F}, the surface density estimates are effectively scaled to an assumed opacity at 250 \micron\ of 21.6 cm$^{2}$ g$^{-1}$. That opacity model assumed grain growth and ice accumulation for \eten6\ yr at a density, $ n = \eten6$ \cmv\
% OH94
\citep{1994A&A...291..943O}, implausible conditions for most molecular cloud material. 
The most recent evaluation of this opacity for molecular clouds (KP5) is 8.12 cm$^{2}$ g$^{-1}$
% Pontoppidan 25
\citep{2024RNAAS...8...68P}, constructed to reproduce opacity laws in molecular clouds by
%Chapman et al. 
\citet{2009ApJ...690..496C}
and quite consistent with recent JWST observations
% Rubinstein 24
\citep[e.g.,][]{2024ApJ...974..112R}.
We use that opacity instead of the one used by
\citet{2025MNRAS.538.2744F}.

\begin{figure*}[ht!]
\includegraphics[width=0.33\textwidth]{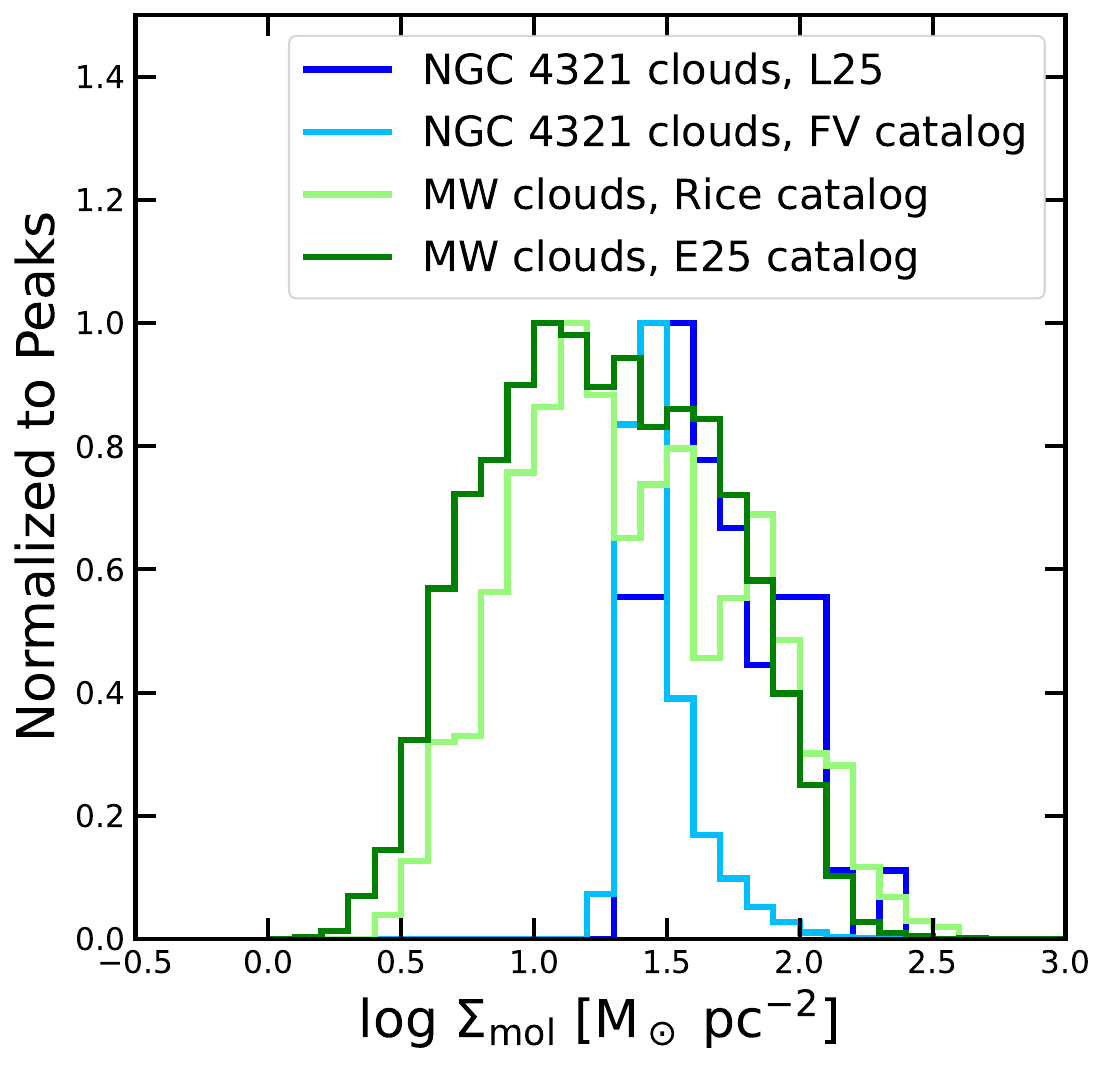}
\includegraphics[width=0.33\textwidth]{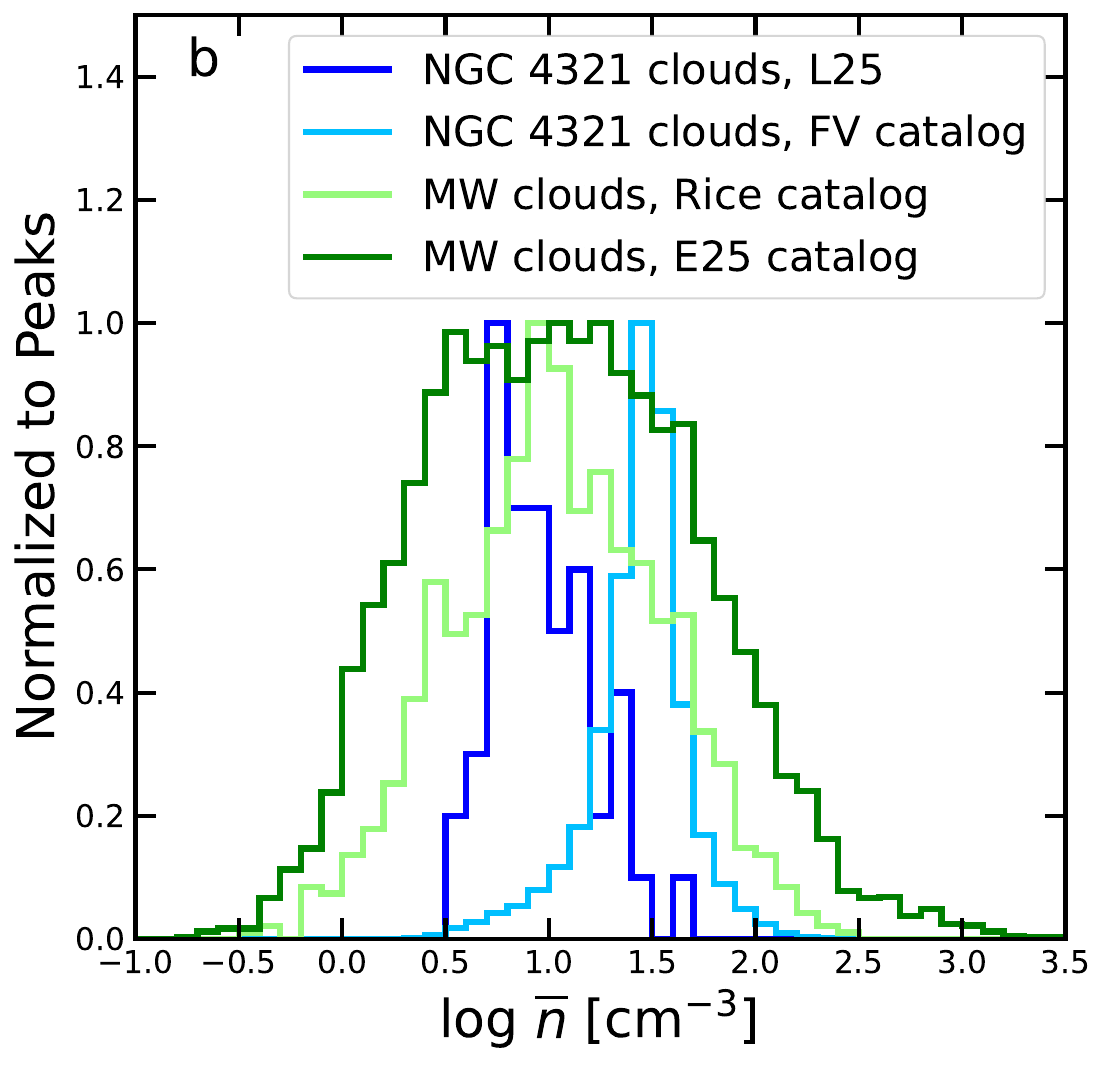}
\includegraphics[width=0.33\textwidth]{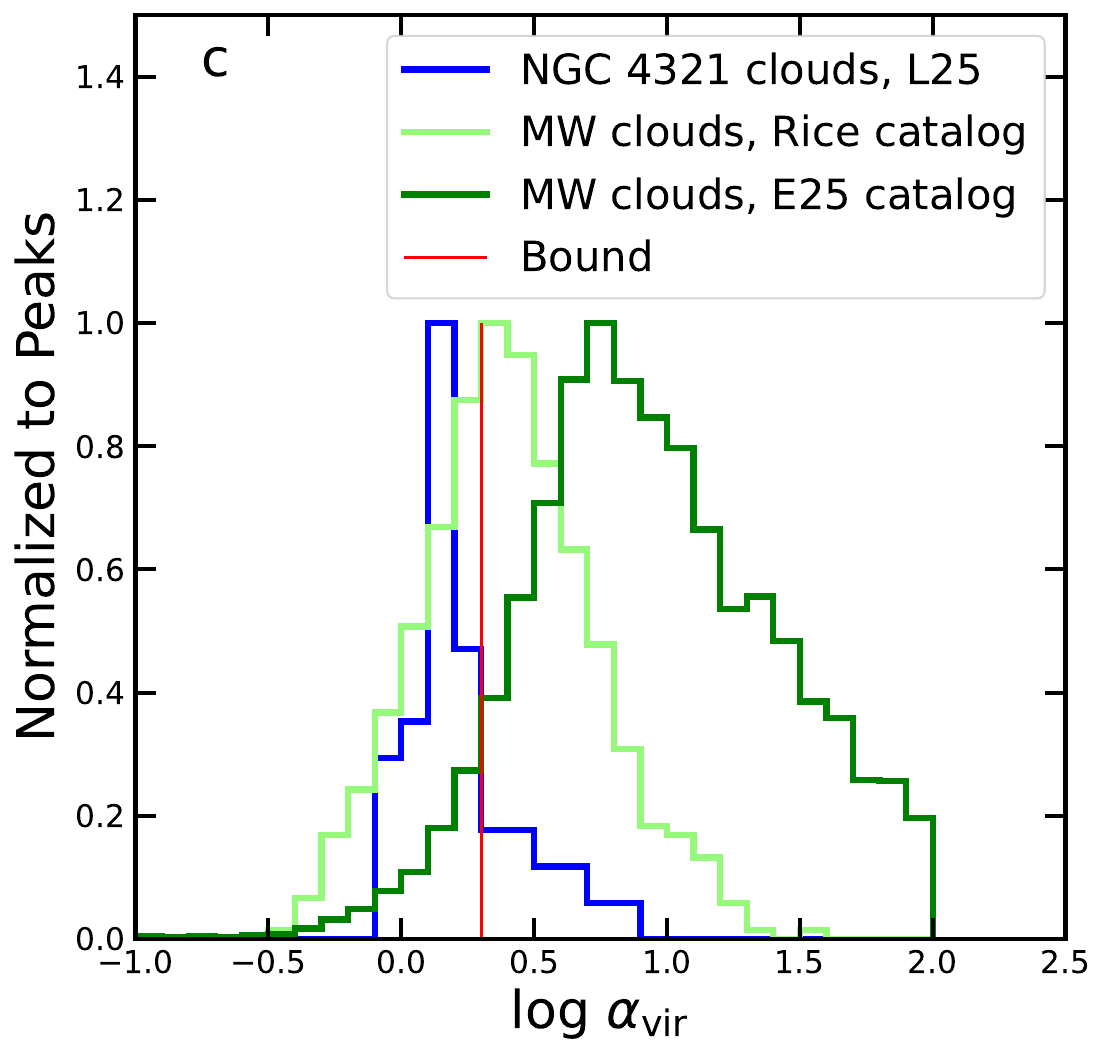}
\caption{The histograms of (a) surface density (\sigmam), (b) average density (\nbar), and  (c) virial parameter (\alphavir) for the \mw\ (from 
%E25
\citetalias{2025ApJ...980..216E} in green and from \citetalias{2016ApJ...822...52R} in light green, respectively) and for NGC4321, both from CO \jj21\ \citepalias{2025ApJ...985...14L} in blue and extinction \citep{2025MNRAS.538.2744F} in deep sky blue, respectively. The last distribution has been scaled to account for lower opacities as discussed in \S \ref{sec:tracers}.}
\label{fig:surfdens}
\end{figure*}

For the \mw, we use the catalog in 
%E25
\citetalias{2025ApJ...980..216E}, but also the catalog from 
%Rice
\citet{2016ApJ...822...52R}, labeled Rice in figures and tables; this catalog was made from the same CO \jj10\ data but a different cloud identification method. 
 We are using here the original cloud catalogs, not values averaged into hexagons.
In particular we compare surface density (\sigmam, Figure \ref{fig:surfdens}), mean density (\nbar, Figure \ref{fig:surfdens}),  and virial parameter (\alphavir, Figure \ref{fig:surfdens}). Statistics of the distributions are presented in Table \ref{tab:tracer}.

%----------Table Tracers and Definitions----------   
\begin{deluxetable*}{l r r r r r r r r r r r r } 
\tablecaption{Effects of Tracers and Definitions \label{tab:tracer}} 
\tablewidth{0pt} 
\tablehead{   
\colhead{Sample } &  \colhead{Number}   & \multicolumn{3}{c}{$\log \nbar$ (\cmv)}   & \multicolumn{3}{c}{\sigmam\ (\msunpc)}   & \multicolumn{3}{c}{\alphavir }   & \colhead{$\log M_{\rm tot}$ } & \colhead{Note} \\  
 & & Med & Mean & Std. & Med & Mean & Std.  &     Med & Mean & Std.          &  (\msun)       &  
  } 
\startdata 
\hline 
NGC4321 \av\  & 24196 & 1.46 & 1.43 & 0.24 & 27.3 & 31.4 & 16.3 & \nodata & \nodata & \nodata & 9.36 & 1 \cr 
NGC4321 \jj21\  & 124 & 0.92 & 0.96 & 0.24 & 48.3 & 61.5 & 41.5 & 1.51 & 1.95 & 1.24 & 9.64 & 2 \cr 
MW Rice & 1037 & 1.04 & 1.04 & 0.50 & 21.2 & 38.3 & 43.4 & 2.44 & 3.45 & 3.34 & 8.39 & 3 \cr 
MW E25 & 7389 & 1.05 & 1.07 & 0.65 & 17.9 & 28.7 & 29.4 & 8.03 & 15.41 & 18.40 & 9.13 & 4 \cr 
\enddata  
\tablecomments{ 1. Data from catalog in  \citet{2025MNRAS.538.2744F} using extinction to define clouds. \\ 2.  Data from catalog in \citet{2025ApJ...985...14L} using CO \jj21\ to define clouds. \\ 
  3. Data from catalog in \citet{2016ApJ...822...52R}, using higher threshold to define clouds. \\ 
 4. Data from catalog in  \citet{2025ApJ...980..216E} using criteria of E25: $\mmol > 1$ \msun, $R_{\rm gal} < 20$ kpc,  and $\alphavir < 100.$ \\  } 
 \end{deluxetable*}

Tracer choice can have a substantial effect on derived surface density (panel a of Figure \ref{fig:surfdens}). The distribution from extinction mapping \citep{2025MNRAS.538.2744F} finds surface densities lower by about a factor of two compared to the CO \jj21\ distribution from the same galaxy. (Before adjusting the opacities, the factor was 5.2.)
This difference is probably mostly due to the lower surface density sensitivity limit for the extinction method and the mass-weighting for the CO data, as well as limitations on the maximum surface densities measurable with the extinction method
\citep{2023MNRAS.524..161F,2025MNRAS.538.2744F}.
The extinction values are however similar to \mw\ values.

The two methods of cloud identification for \mw\ clouds have quite similar distributions and statistical properties, though the 
Rice values are 1.2-1.3 times higher. Tracer choice seems to have a larger effect on surface density estimates than does cloud identification method, especially with uncertainties in opacity laws.

For the volume density (\nbar), the results are quite different (panel b of Figure \ref{fig:surfdens}). While the four distributions differ in width, their medians and means are striking in their similarity. The smaller cloud sizes for the extinction method result in somewhat higher \nbar\ values, but still only different by a factor of three.

%\begin{figure}[h!]
%\includegraphics[width=0.47\textwidth]{densities.pdf}
%\caption{The histograms of mean cloud density (\nbar) for the \mw\ (from 
%E25
%\citetalias{2025ApJ...980..216E} in green and from 
%Rice 
%\citetalias{2016ApJ...822...52R} in light green) and for NGC4321, both from CO~\jj21\ \citepalias{2025ApJ...985...14L} in blue and extinction \citep{2025MNRAS.538.2744F} in sky blue, respectively.The last distribution has been scaled to account for lower opacities as discussed in \S \ref{sec:tracers}.}
%\label{fig:densities}
%\end{figure}

The distributions of virial parameter present yet a different story (panel c of Figure \ref{fig:surfdens}). Extinction measurements provide no measure of virial parameter, so we have only three distributions. The CO \jj21\ distribution is centered at lower values of \alphavir\ than are the \mw\ distributions, though the Rice distributions are much more similar than those of 
E25. This difference has been discussed by 
% Evans 21
\citet{2021ApJ...920..126E}. 

%\begin{figure}[h!]
%\includegraphics[width=0.47\textwidth]{virial.pdf}
%\caption{The histograms of virial parameter (\alphavir) for the \mw\ (from 
%E25
%\citetalias{2025ApJ...980..216E} in green and from 
%R16
%\citetalias{2016ApJ...822...52R} in light green) and for NGC4321 from CO \jj21\ %\citepalias{2025ApJ...985...14L} in blue. Clouds to the left of the vertical red line are gravitationally bound. 
%The cutoff at $\log \alphavir = 2.0$ in the 
%E25
%\citetalias{2025ApJ...985...14L} distribution is due to the selection criterion.}
%\label{fig:virial}
%\end{figure}

To summarize, the choice of tracer has a direct effect on estimates of surface density, but little apparent effect on volume density for the three considered here. In contrast, the cloud identification method has a major effect on virial parameter, but modest effects on surface density and volume density.

\section{Molecular gas depletion time as a function of other properties}\label{sec:tdep}

In this section, we compare the \mw\ to results from 
PHANGS with plots like those in \S 3.1 to \S 3.3 of 
%L25
\citetalias{2025ApJ...985...14L}.  Table \ref{tab:statskey} provides the statistical properties of the variables for the \mw, while Table \ref{tab:statskeyL25} contains the equivalent statistics for the data in the Erratum, \citep{2026ApJ..1003..248L}.
We follow the procedures in \citetalias{2025ApJ...985...14L} as closely as possible. In particular, we do not apply any criteria on \ico\ or \sigmam\ at the cloud level; the criterion on \sigmam\ is applied only to analysis of data at the hexagon level, consistent with step 6 in \S \ref{sec.proc}. Therefore the statistical properties of the variables in Table \ref{tab:statskey} differ from those in Table \ref{tab:res}. 
As explained in \citet{2026ApJ..1003..248L}, some points in the figures published by \citetalias{2025ApJ...985...14L} were incorrectly placed. We show the equivalent to Table~3 in \citetalias{2025ApJ...985...14L} with correlation and fit information in Table \ref{tab:correlations2}. 

%----------Table Statisics of Key parameters-Default-----------------------   
\begin{deluxetable}{l r r r l } 
\tablecaption{Statistics of Key Parameters For \mw \label{tab:statskey}} 
\tablewidth{0pt} 
\tablehead{\colhead{Quantity} & \colhead{Median} &  \colhead{Mean} & \colhead{Std Dev} & \colhead{Units} }  
\startdata 
\hline 
$\tdephex$  & $1.46$ & $3.76$ & $9.98$ & Gyr \\ 
$\tffhex $ & $9.34$ & $9.23$ & $2.50$ & Myr \\ 
\sigsmhex  & $42.59$ & $58.98$ & $49.56$ & \msunpc \\ 
\sigvcellhex  & $5.71$ & $5.94$ & $1.82$ & \kms \\ 
\avirhex  & $3.94$ & $4.85$ & $3.14$ & \nodata \\ 
\epsffhex  & $0.57$ & $0.89$ & $0.98$ & \% \\ 
\epsffEKO  & $0.55$ & $0.95$ & $1.30$ & \% \\ 
\enddata  
\tablecomments{  1. \epsffhex\ is \tffhex/\tdephex\ for each hexagon. \\ 
 2. \epsffEKO\ is calculated for each hexagon from \avirhex\ and the EKO formula. \\ 
 } 
 \end{deluxetable}

The statistics in Table \ref{tab:statskey} would apply to the \mw, as analyzed as much as possible as a galaxy in the PHANGS sample. Comparing the mean values to the entries in Table \ref{tab:statskeyL25} for the \citetalias{2025ApJ...985...14L} sample indicates that the depletion and free-fall times for the \mw\ are  longer, the velocity dispersion is lower, the virial parameter is higher, and the surface densities are very similar. All these properties have large dispersions, so the \mw\ clearly fits within the distribution of properties of the \citetalias{2025ApJ...985...14L} sample. The efficiency per free-fall time, averaged over all hexagons, has a mean value of 0.91\% in the \mw\ and 0.55\% for \citetalias{2025ApJ...985...14L}. The ``theoretical" version, \epsffEKO,  calculated from the virial parameters of each hexagon, agrees with the observed value in the \mw, but is nearly 4 times larger than the observed value in the \citetalias{2025ApJ...985...14L} sample.

%----------Table Statisics of Key parameters for L25 -Default-----------------------   
\begin{deluxetable}{l r r r l } 
\tablecaption{Statistics of Key Parameters For L25 Data\label{tab:statskeyL25}} 
\tablewidth{0pt} 
\tablehead{\colhead{Quantity} & \colhead{Median} &  \colhead{Mean} & \colhead{Std Dev} & \colhead{Units} }  
\startdata 
\hline 
$\tdep$  & $2.13$ & $2.50$ & $1.76$ & Gyr \\ 
$\tffpm $ & $8.68$ & $8.62$ & $2.40$ & Myr \\ 
\mean{\sigmolcloud}  & $38.14$ & $56.52$ & $54.71$ & \msunpc \\ 
\sigvpm  & $5.43$ & $7.12$ & $4.95$ & \kms \\ 
\avirpm  & $2.01$ & $3.08$ & $3.95$ & \nodata \\ 
\epsff(hex)  & $0.39$ & $0.55$ & $0.49$ & \% \\ 
\epsffEKO  & $1.72$ & $1.90$ & $1.31$ & \% \\ 
\enddata  
\tablecomments{  1. \epsff (hex) is \tff/\tdep\ for each hexagon. \\ 
 2. \epsffEKO\ is calculated for each hexagon from \avirhex\ and the EKO formula. \\ 
 } 
 \end{deluxetable}

Following \citetalias{2025ApJ...985...14L}, we test correlations with the Spearman test, with $r_s$ the correlation coefficient and $p$ the probability that the assumption of no correlation can be rejected. Values of $p \leq 0.05$ are needed for rejection of the null hypothesis. We also fit the data in each plot to the formula
\begin{equation}\label{eq:fiteq}
 y = m(x-x_0) + b,
\end{equation}
where $y$ is $\log_{\rm 10} \tdep$ and $x$ is the decimal logarithm of the variable being tested, while $x_0$ is the median of $x$.
We use a different fitter (LevMarlsq from astropy modeling) but our results differ from those in \citet{2026ApJ..1003..248L} by only insignificant amounts.

%\input tabcorr.txt

%----------Table correlations with Uncertainties-----------------------   
\begin{deluxetable*}{l l r r r r r r r} 
\tablecaption{Correlation coefficients and fit parameters of log \tdep\ as function of $x$ \label{tab:correlations2}} 
\tablewidth{0pt} 
\tablehead{\colhead{Sample} & \colhead{$x$} &  \colhead{$r_{\rm s}$} & \colhead{$p$} & \colhead{$m$}  & \colhead{$\sigma_{\rm m}$} & \colhead{$x_0$}   & \colhead{$b$}   & \colhead{$\sigma_{\rm b}$} }  
\startdata 
\hline 
PHANGS & log \tffhex & $0.18$ & 2.81e-07 & $0.51$ & 0.08 & 6.94 & 9.31 & 1.06e-02  \\ 
MW  & log \tffhex & $-0.04$ & 6.43e-01 & $-0.10$ & 0.28 & 6.97 & 9.23 & 3.79e-02  \\ 
MW (PHANGS) & log \tffhex & $-0.04$ & 6.43e-01 & $-0.20$ & 0.25 & 6.97 & 9.17 & 3.79e-02  \\ 
PHANGS  & log \sigsmhex & $-0.22$ & 1.40e-10 & $-0.31$ & 0.04 & 1.58 & 9.32 & 1.07e-02  \\ 
MW  & log \sigsmhex & $0.04$ & 5.82e-01 & $0.06$ & 0.14 & 1.63 & 9.22 & 3.80e-02  \\ 
MW (PHANGS) & log \sigsmhex & $0.04$ & 5.82e-01 & $0.10$ & 0.13 & 1.63 & 9.17 & 3.80e-02  \\ 
PHANGS  & log \sigvcellhex  & $-0.31$ & 6.28e-20 & $-0.51$ & 0.05 & 0.74 & 9.33 & 1.05e-02  \\ 
MW  & log \sigvcellhex  & $-0.29$ & 2.34e-04 & $-1.15$ & 0.26 & 0.76 & 9.22 & 3.53e-02  \\ 
MW (PHANGS) & log \sigvcellhex  & $-0.29$ & 2.34e-04 & $-0.71$ & 0.28 & 0.76 & 9.16 & 3.53e-02  \\ 
PHANGS  & log \avirhex & $-0.27$ & 3.67e-15 & $-0.32$ & 0.03 & 0.30 & 9.31 & 3.41e-02  \\ 
MW  & log \avirhex & $-0.28$ & 2.86e-04 & $-0.39$ & 0.12 & 0.60 & 9.23 & 3.62e-02  \\ 
MW (PHANGS) & log \avirhex & $-0.28$ & 2.86e-04 & $-0.31$ & 0.11 & 0.60 & 9.17 & 3.62e-02  \\ 
\enddata  
\tablecomments{ 1. MW (PHANGS) denotes MW data restricted to range of PHANGS data. \\  } 
 \end{deluxetable*}

First, we reproduced Figure~4 through~7 in \citetalias{2025ApJ...985...14L} (1 through~4 in \citet{2026ApJ..1003..248L}).
After checking that the corrected plots were reproduced, we retained only a fit to the \citetalias{2025ApJ...985...14L} data as a blue line for comparison in the plots of \mw\ data. The data points for the \mw\ are shown as filled circles, color-coded by \rgal.   The fit to the \mw\ data is shown as a green line. For each plot, we also fitted the data restricted to the region occupied by the data in \citetalias{2025ApJ...985...14L}, indicated by the orange box, and those fits are represented by orange lines. The right axis shows the data normalized to the median value of depletion time for comparison to those figures in \citetalias{2025ApJ...985...14L}.

\begin{figure*}[ht!]
\includegraphics[width= 0.50\textwidth]{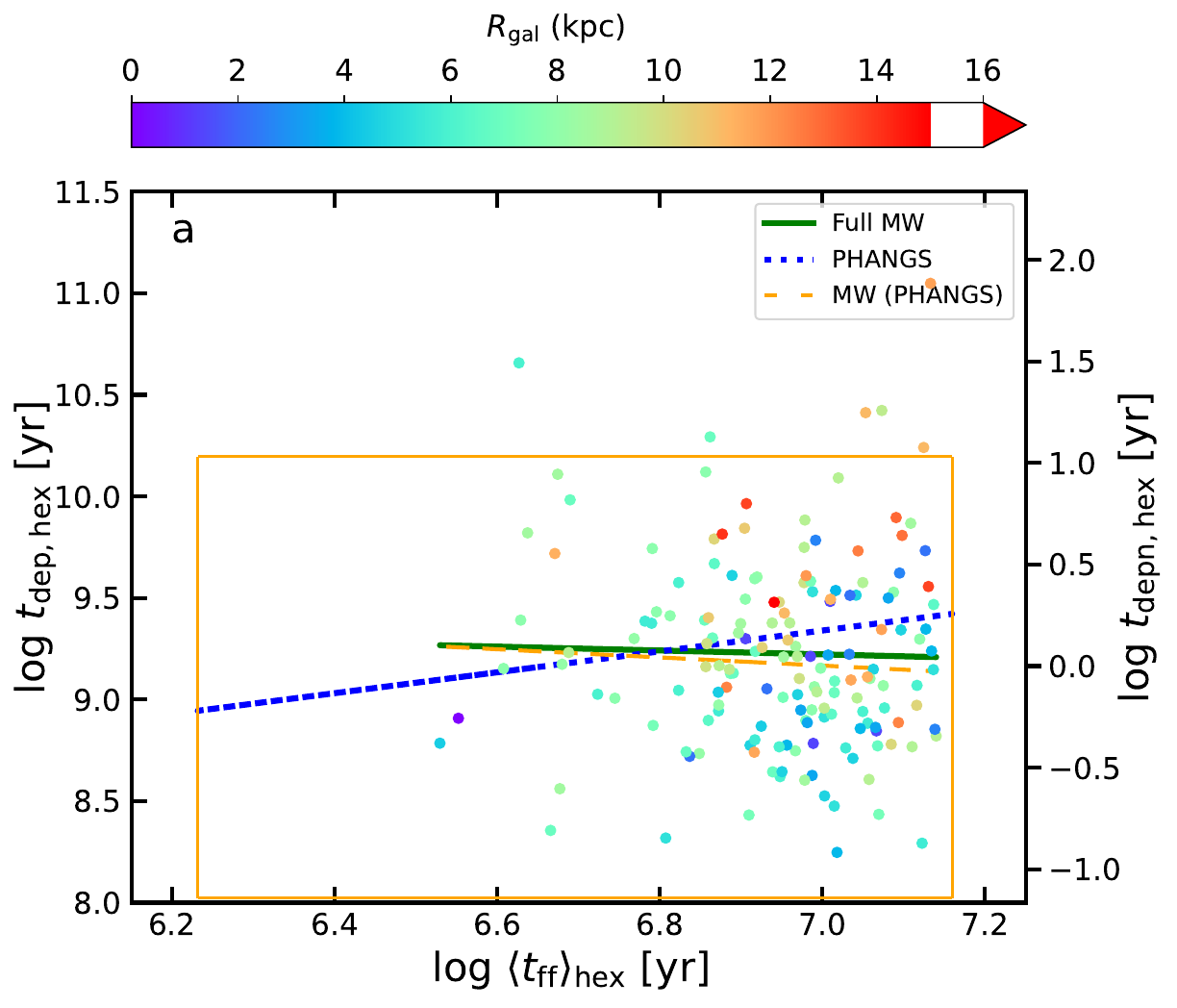}
\includegraphics[width= 0.50\textwidth]{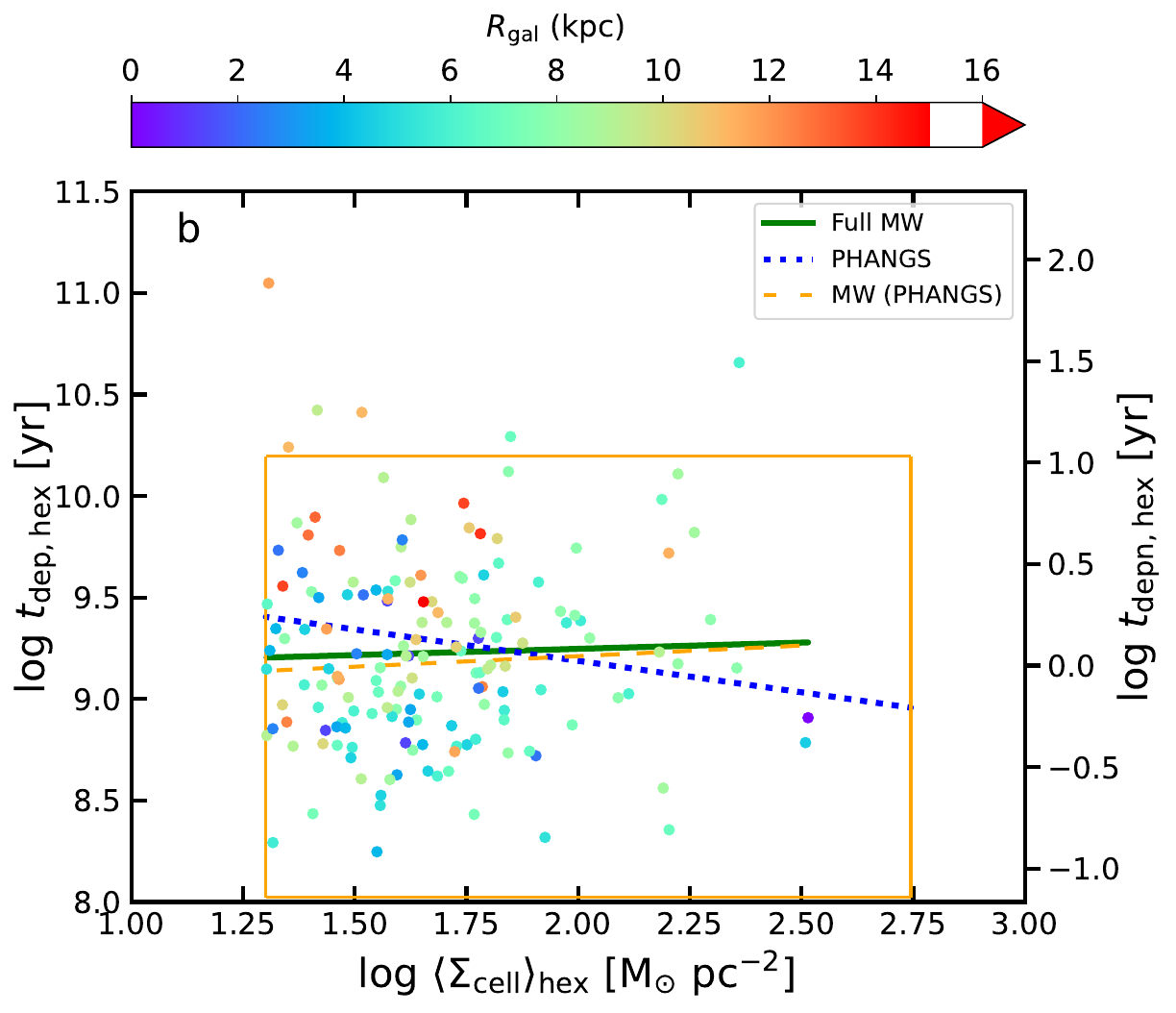}
\caption{In panel a, the hex-level depletion time is plotted versus cloud-average free-fall time, mass-weighted averaged into hexagons, for the \mw. The covariance of the uncertainties in these parameters is shown in Figure~4 of \citetalias{2025ApJ...985...14L}. 
Panel b has the equivalent plot for \sigsmhex. In both panels,
the points are color-coded to show \rgal. The rainbow color scale is capped at 15 kpc for better clarity. The fit to the \mw\ data is shown in green, and the fit to \mw\ values restricted to the range of those in \citetalias{2025ApJ...985...14L} is shown in orange. The orange box is the region including all the points in the PHANGS data.
The blue line is the fit to the equivalent data on the PHANGS galaxies.
The right axis shows the data normalized to the median value of depletion time for comparison to those figures in \citetalias{2025ApJ...985...14L}.
}
\label{fig:ALf4-tff-MW-alt}
\end{figure*}

Figure \ref{fig:ALf4-tff-MW-alt} (the equivalent of Figure~4 of \citetalias{2025ApJ...985...14L}, Figure~1 of \citet{2026ApJ..1003..248L}) shows no significant correlation between the depletion time and the free-fall time, 
using our hex-level data. 
Including only the points lying within the boundaries of the \citetalias{2025ApJ...985...14L} data removes the hexagons with large \tdep, but the fit is scarcely affected. The \mw\ lacks hexagons with very low values of free-fall time found in the PHANGS sample, albeit rarely, as shown by the empty area to the left of the orange box.

Next we make plots analogous to their figures~6 and~7 of \citetalias{2025ApJ...985...14L} (Figure 3 of the Erratum), corresponding to the fits in their Table~3. 
The right panel of figure~\ref{fig:ALf4-tff-MW-alt} shows the depletion time versus mass surface density.
For the \mw, we use the surface density, smoothed over the cell and mass-weighted averaged into hexagons, \sigsmhex. This is closest in spirit to what \citetalias{2025ApJ...985...14L} does.
We see a barely significant correlation.

\begin{figure*}[ht!]
\includegraphics[width= 0.50\textwidth]{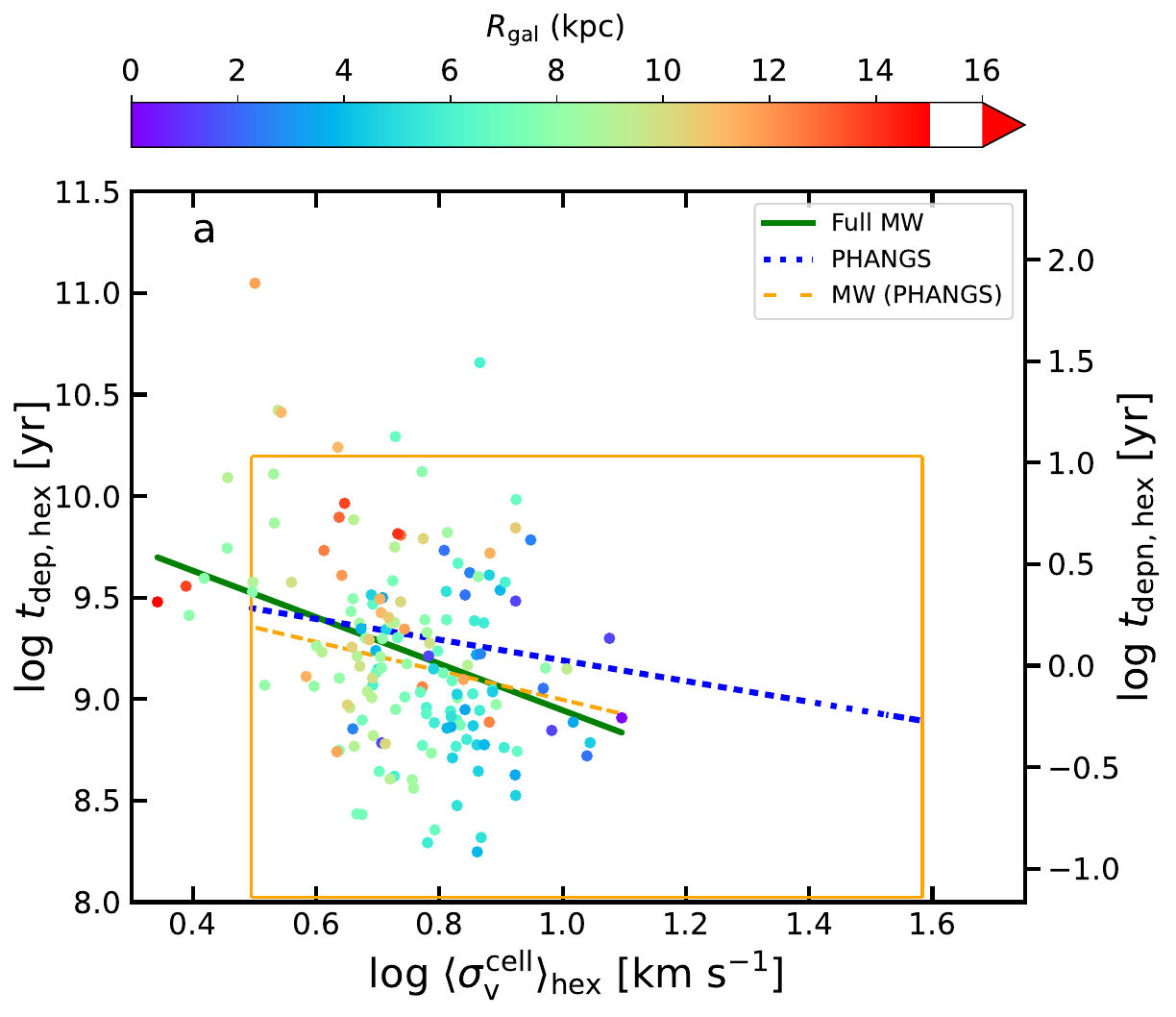}
\includegraphics[width= 0.50\textwidth]{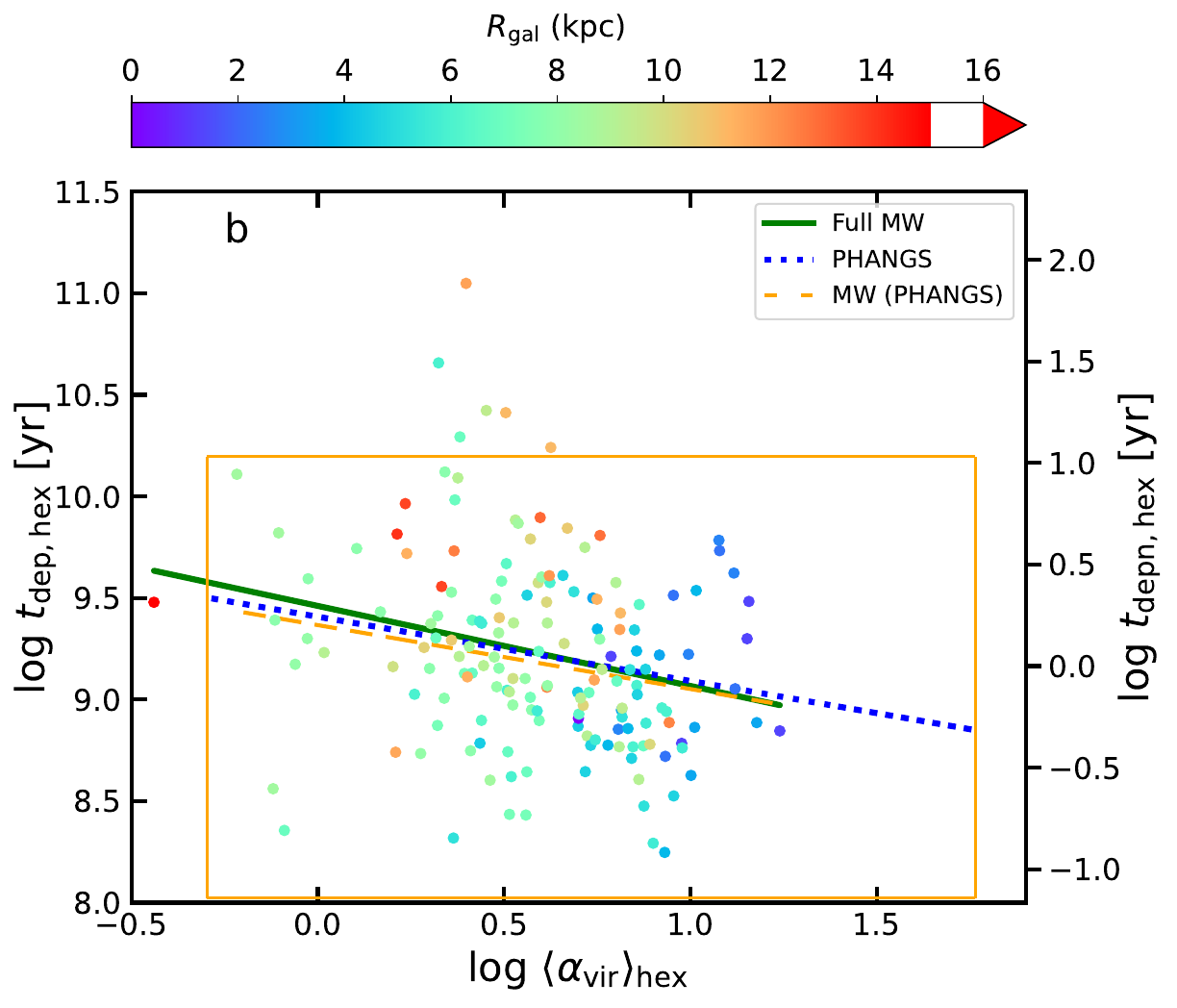}
\caption{In panel a, the mean depletion time is plotted versus the cloud-average velocity dispersion, mass-weighted averaged into hexagons, for the \mw. In panel b, the plot is versus \avirhex. For both panels, the points are color-coded to show \rgal. The rainbow color scale is capped at 15 kpc for better clarity. The fit to the \mw\ data is shown in green, and the fit to \mw\ values restricted to the range of those in \citetalias{2025ApJ...985...14L} is shown in orange. The orange box is the region including all the points in the PHANGS data.
The blue line is the fit to the equivalent data on the PHANGS galaxies.
The right axis shows the data normalized to the median value of depletion time for comparison to those figures in \citetalias{2025ApJ...985...14L}.
}
\label{fig:ALf6-sigv-MW-alt}
\end{figure*}

The average $\sigmav$ over the clouds in a cell, mass averaged into hexagons \citepalias[panel a of Figure~\ref{fig:ALf6-sigv-MW-alt}, equivalent to Figure~6 of][the lower panel of Figure~3 of \citet{2026ApJ..1003..248L}]{2025ApJ...985...14L}
shows, like the extragalactic data, a negative correlation, but the \mw\ plot lacks the large \sigmav\ points. At least in part, this is because the \mw\ values are average \sigmav\ at the cloud level, which do not include velocity differences between clouds that are combined into \sigmav\ with the spatial resolution of 150~pc.

The dependence of depletion time on virial parameter \citepalias[panel b of Figure \ref{fig:ALf6-sigv-MW-alt}, equivalent to Figure~7 of][and~4 of \citep{2026ApJ..1003..248L}]{2025ApJ...985...14L} indicates a slight negative correlation, very similar to that of the PHANGS data. Recall that our star formation rate is based on the 70 \micron\ emission, \sfrseventy. If we use the CO mass and the formula from EKO that is based on a relation with \alphavir, we get a strong positive correlation, but this would be circular.

\begin{figure}[ht!]
\includegraphics[width= 0.47\textwidth]{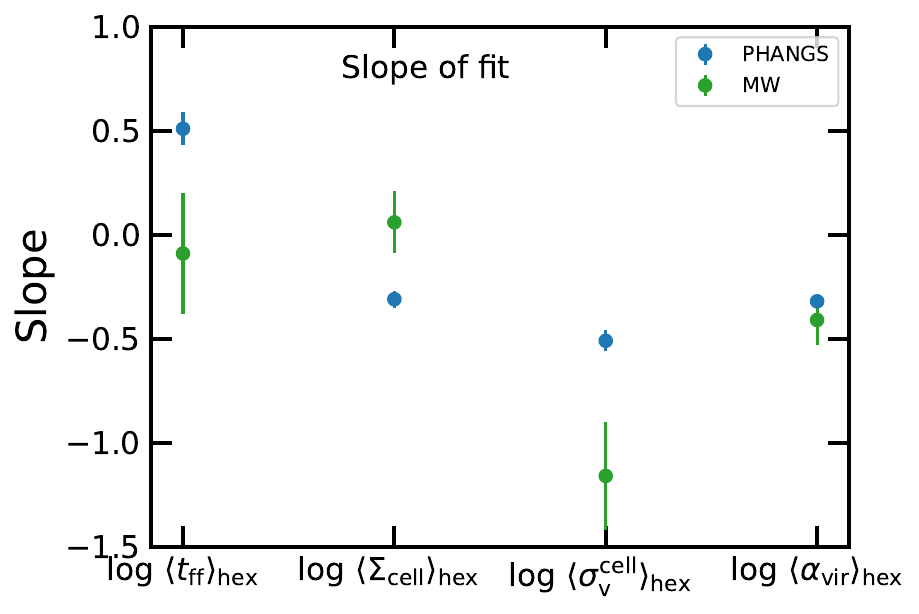}
\caption{The slopes of the fits in Table \ref{tab:correlations2} are plotted with uncertainties for both the PHANGS galaxies and the \mw.
}
\label{fig:compareslopes}
\end{figure}

This section has shown that the \mw, analyzed similarly to the PHANGS sample, has similar dependences of \tdep\ on various parameters. There are relatively strong anti-correlations between \tdep\ and line width or virial parameter. Correlations between \tdep\ and \tffhex\ or \sigsmhex\ are very weak and probably not significant; these were also weak but more significant in the PHANGS data. 
A plot of the slopes of the fitted lines with uncertainties for both the PHANGS galaxies and the \mw\ (Figure \ref{fig:compareslopes}) shows that these differ but only at the 2~$\sigma$ level. The errorbars on the \mw\ are larger because there are many fewer data points. 

As 
%L25
\citetalias{2025ApJ...985...14L} note, the results of these correlations change substantially if a fixed value of \alphaco\ is used instead of the standard procedure. We confirm this fact (Appendix \ref{app:alphaco}, Table \ref{tab:corr2fixeda}) and find that the differences in the fits between the \mw\ and the PHANGS galaxies are smaller with this choice
(Figure \ref{fig:compareslopesfix}).
While we agree with
%L25
\citetalias{2025ApJ...985...14L}
that the evidence for a variable \alphaco\ is strong, remaining uncertainties in this function are problematic.

\section{Radial Distributions}\label{sec:radial}

\begin{figure}[ht!]
\includegraphics[width=0.47\textwidth]{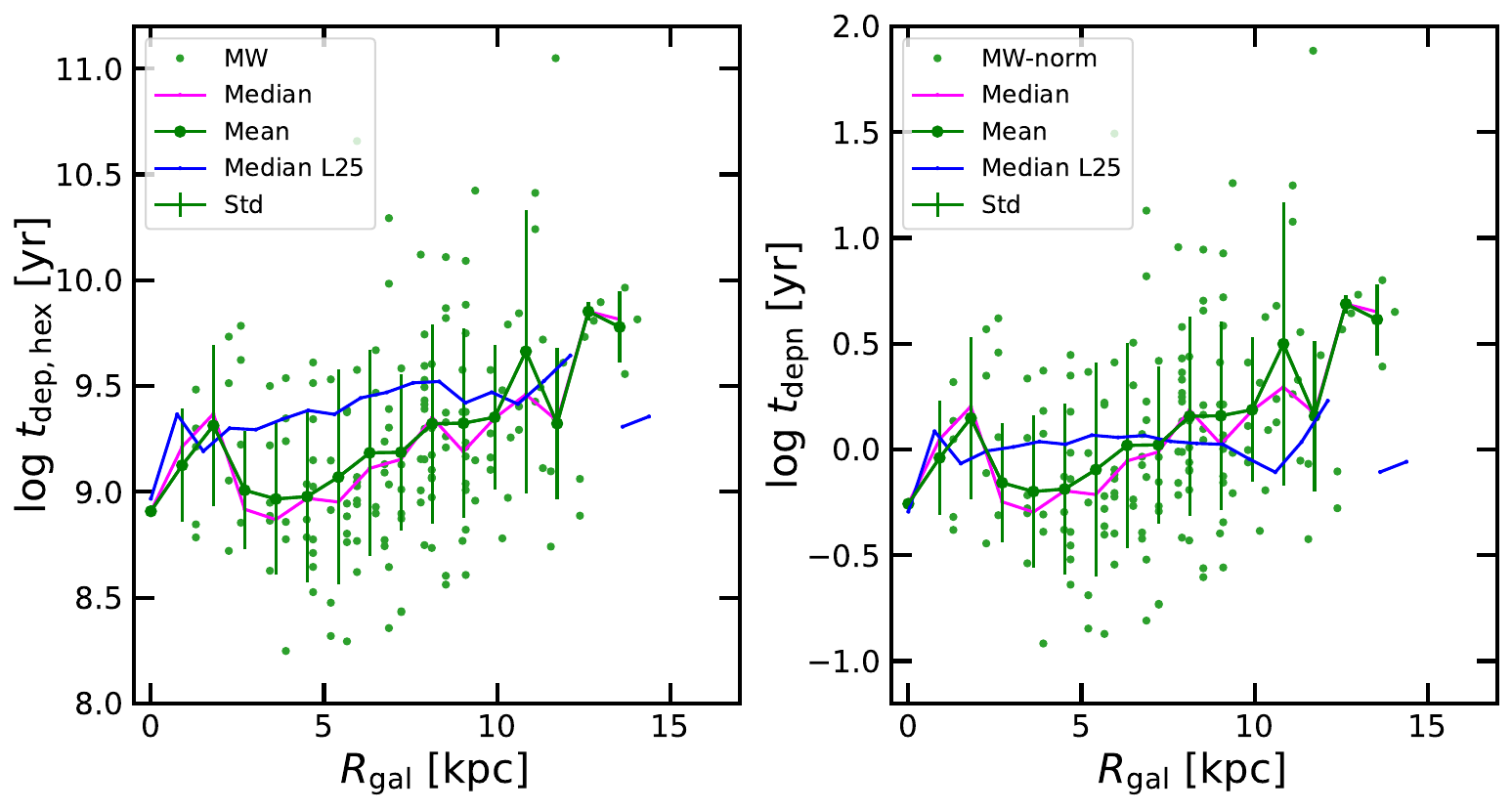}
\includegraphics[width=0.47\textwidth]{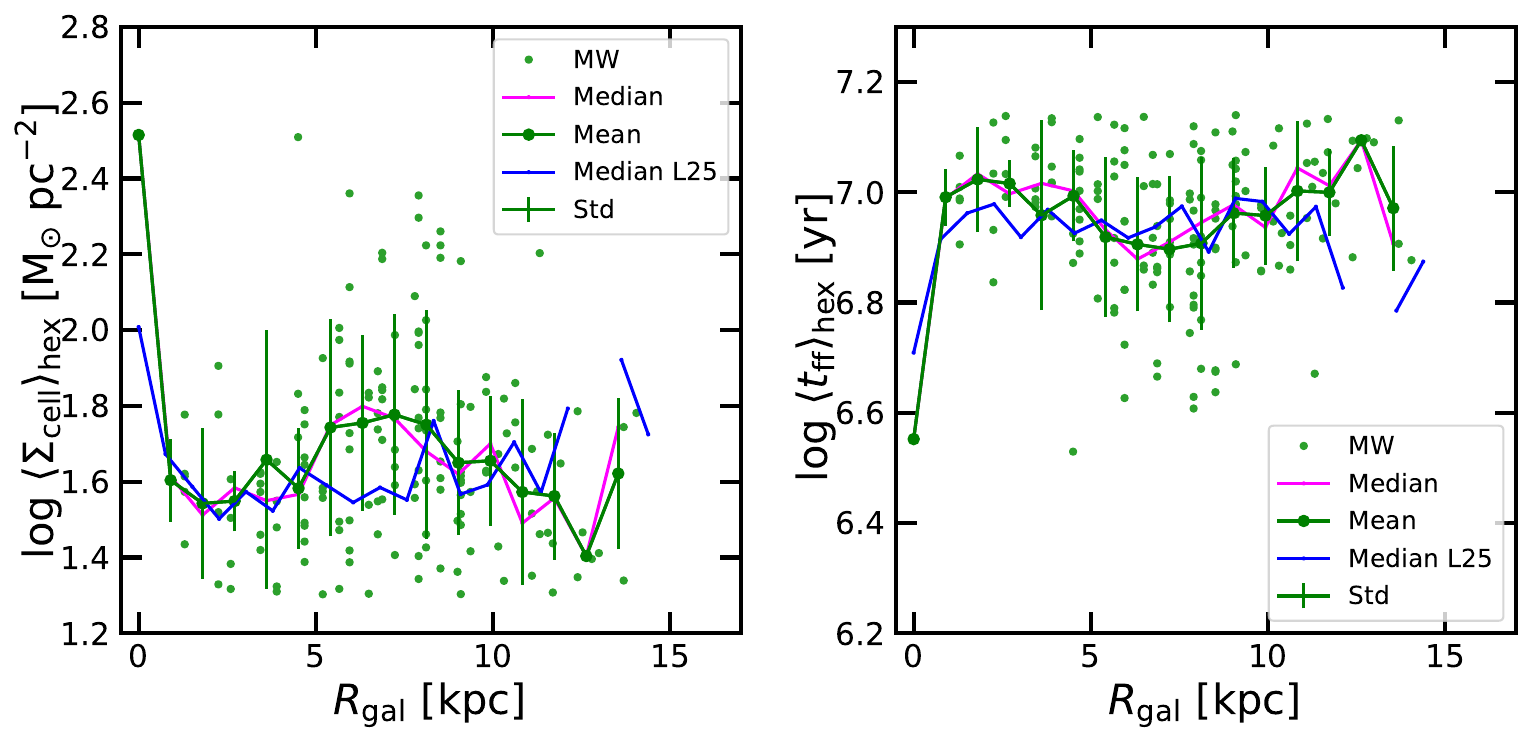}
\includegraphics[width=0.47\textwidth]{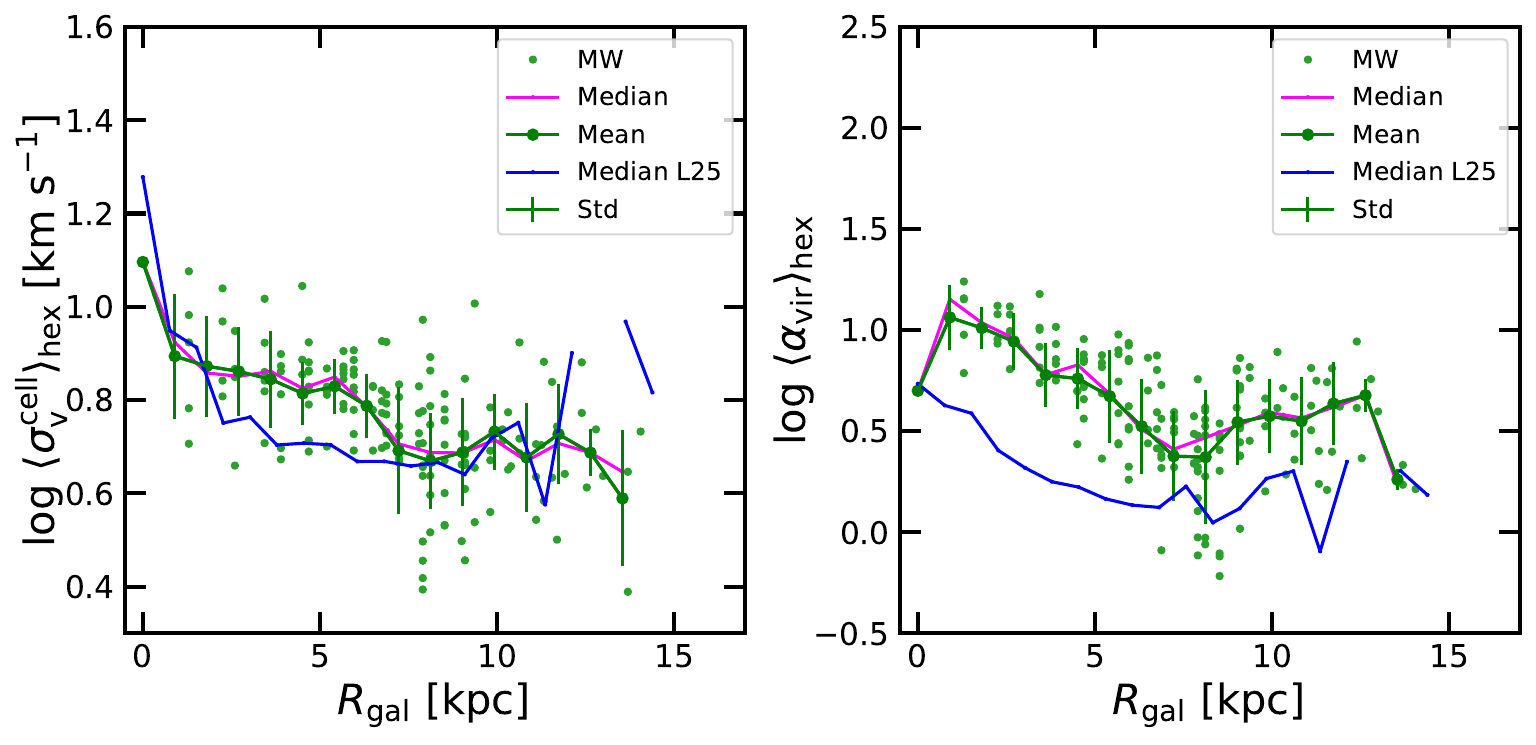}
\caption{\mw\ version of Figure~8 of \citetalias{2025ApJ...985...14L}. The magenta line shows the running median with 20 bins, while the green line with error bars shows the mean and standard deviation. The blue line is the running median of the \citetalias{2025ApJ...985...14L} data. The quantities on the y-axis follow the definitions in Table~\ref{tab:symbols}. }
\label{fig:ALf8-MW}
\end{figure}

Many of the properties we are studying vary systematically with \rgal, as can be seen from the color coding, so we also plotted them versus \rgal, following the procedures in \citetalias{2025ApJ...985...14L}. We first reproduced the plots of properties in the PHANGS sample versus galactocentric radius in \citetalias{2025ApJ...985...14L}. Then we plotted the same properties versus \rgal\ for the \mw.
Figure~\ref{fig:ALf8-MW} is the equivalent of Figure~8 of \citetalias{2025ApJ...985...14L} (Figure~5 in \citet{2026ApJ..1003..248L}) for the \mw. Patterns are similar. Rather than their running median method, we use scipy binned statistics to compute median, mean, and standard deviation in 20 bins. The \mw\ shows very different values in the innermost hexagon (henceforth the CMZ) in most plots, similar to many PHANGS galaxies.
As is true for the PHANGS data, the only variables that show a clear trend  with \rgal\ outside the CMZ are \sigvcellhex\ and \avirhex. 
%MWISP 25
\citet{2025AJ....170..239Z} also see a clear decrease of velocity dispersion with \rgal\ for the clouds in their survey area.

The \mw\ and the galaxies in \citetalias{2025ApJ...985...14L} generally exhibit the same trends with \rgal. However,  in the region of $ 1 < \rgal < 8$ kpc, the \mw\ tends to have higher \sigvcellhex\ and \avirhex. These differences may be related to the fact that the \mw\ has a decrease in the molecular surface density and star formation rate 
% Elia 22, 25
\citep{2022ApJ...941..162E,2025ApJ...980..216E},
associated with a bar in the inner \mw\ 
% Benjamin, Lucey
\citep{2005ApJ...630L.149B, 2023MNRAS.520.4779L}
While other barred galaxies support the interpretation of the \mw\ data as a real dip in the molecular gas
% Evans 25
\citep{2026ApJ...998L..23E}, it is still true that distances in that region may be incorrect. A recent analysis of the \mw\ data using simulations of barred galaxies also found a dip of about the same depth, but perhaps a smaller radial extent 
%Baba 26
\citep{2026PASJ...78.1093B}.

\section{Discussion}\label{sec:disc}

When analyzed in a similar way, the \mw\ data occupies areas in the plots of \tdep\ versus other parameters that are similar to the PHANGS galaxies with some differences. In particular, both datasets show tremendous scatter about any relationships. On the scale of 1.5 kpc hexagons, mass-weighted properties of clouds/cells of diameter 150 pc produce little correlation between depletion time and either free-fall time or mass surface density. The strongest correlation is a negative one between depletion time and velocity dispersion, seen in both the PHANGS sample and the \mw, suggesting faster star formation in more turbulent gas. Velocity dispersion correlates with mass, and more massive clouds are more likely to be bound
% Evans 21
\citep{2021ApJ...920..126E}, possibly explaining the correlation. On the other hand, \tdep\ anticorrelates with virial parameter, suggesting faster star formation in {\it less} bound gas. This fact, seen in \citetalias{2025ApJ...985...14L}, is reproduced in the \mw.  We first address the (small) differences between the \citetalias{2025ApJ...985...14L} and \mw\ results. Then we discuss expected relations based on theory or empirical arguments. After a digression to consider further the effects of resolution, we discuss the comparison of the observations with expectations.

\subsection{Differences between PHANGS galaxies and the Milky Way}\label{sec:PvsMW}

As seen in Figure \ref{fig:compareslopes}, the analysis of the PHANGS sample and \mw\ differ significantly about the slope in the \tdephex--\tffhex\ diagram. Figure 1 of \citet{2026ApJ..1003..248L} contains a number of data points with \tffhex\ significantly lower than any hexagons in the \mw. If the lower limit on \tffhex\ included in the fit to the L25 data is increased gradually, the correlation disappears and the fitted slope becomes close to zero. The difference between \citetalias{2025ApJ...985...14L} and the \mw\ is largely driven by the lack of hexagons with small \tffhex\ in the \mw. These low values occur most commonly at small galactocentric radii (Figure \ref{fig:ALf8-MW}). Many of these exist in the PHANGS sample, but the \mw\ has only one central region. Similar considerations may explain the difference in slopes in the \tdep--\sigsmhex\ diagram (Figure \ref{fig:ALf4-tff-MW-alt}).

The \mw\ shows a steeper slope in the \tdep--\sigvcellhex\ diagram, perhaps caused by the hexagons with very large \tdep\ and low \sigvcellhex. When the fit is restricted to the area in the \citetalias{2025ApJ...985...14L} plots, the fit flattens to a slope of $-0.61$, more similar to the \citetalias{2025ApJ...985...14L} fit. As noted above the velocity dispersions may be underestimated because they ignore cloud-to-cloud variations that are unresolved in the extragalactic data. We tested the effects of spatial resolution by producing the same plot using the data with 60, 90, and 120 pc resolution in the tables in \citetalias{2025ApJ...985...14L}. The best-fitting slopes hardly differed; for example, for the 60 pc data, the slope was $-0.56 \pm 0.10$, not significantly different from that for 150 pc ($-0.51 \pm 0.05$). Ongoing observations with better spatial resolution will allow further study of this issue.

In summary, the \mw\ differs from the PHANGS galaxies in relatively small ways, and some of the difference may be caused by selection effects.

\subsection{Expected Relations}\label{sec:theory}

As discussed by \citetalias{2025ApJ...985...14L}, simple considerations and relations found for whole galaxies lead to some expectations for the relations plotted in \S \ref{sec:tdep}.
If the free-fall time is the pacing timescale for star formation, we would expect that $\tdep \propto \tff$ if the efficiency per free-fall time (\epsff) is constant
% Krumholz 2012, Pokhrel21
\citep{2012ApJ...745...69K,2021ApJ...912L..19P}. 

On the scale of whole galaxies, \sigmasfr\ correlates strongly with \sigmam. 
As discussed by \citetalias{2025ApJ...985...14L}, clouds with similar \epsff\ should show a dependence of
\begin{equation}\label{eq:tdep-sigm}
\tdep \propto \mean{\sigmam}^{-a}
\end{equation}
with $a = 0.5$.
This is equivalent to a Kennicutt-Schmidt (KS) relation
\begin{equation}\label{eq:KS}
\sigmasfr \propto \mean{\sigmam}^{1+a}.
\end{equation}
The KS relation for total gas is fitted well with $a = 0.4$
%Kennicutt98 KE12
\citep{1998ApJ...498..541K,2012ARA&A..50..531K}, 
while that including only the molecular component favors $a = 0$ 
%Bigiel
\citep{2011ApJ...730L..13B}.
Studies of nearby molecular clouds where star formation and surface density can be traced to small scales find mixed results for $a$. Some studies found a linear dependence of \sigmasfr\ on \sigmam\ ($a = 0$) above a threshold around 120 \msunpc\
%Heiderman 2010 and Lada 2010, 2012
\citep[e.g.][]{2010ApJ...723.1019H,2010ApJ...724..687L,2012ApJ...745..190L}.
More recent studies
that trace the structure within clouds, both in young stars and molecular gas, primarily at $\sigmam > 100$ \msunpc\ find a stronger dependence of \sigmasfr\ on \sigmam, with $a = 1.0$, but a possible decrease in $a$ at very high \sigmam\
% Pokhrel 21
\citep{2021ApJ...912L..19P}. 

If, instead of \tff, the crossing time is the pacing time for star formation
%Elmegreen 2000
\citep{2000ApJ...530..277E}, 
one might expect for fixed size that $\tdep \propto \sigmav^{-1}$; \citetalias{2025ApJ...985...14L} provide a more detailed rationale for this dependence based on the ``Heyer-Keto'' relation
% Heyer 09
\citep{2009ApJ...699.1092H}
and a fixed virial parameter. In reality, neither size nor virial parameter is fixed at the cloud scale. However, size is fixed at 150 pc for the cell-based values and their hexagonal averages. Thus the crossing time for a region on the scale of 150 pc may be a relevant timescale.

Simulations predict that the star formation rate decreases rapidly with increasing virial parameter because the efficiency per free-fall time depends exponentially on the square root of  \alphavir\
%Padoan, Kim
\citep{2012ApJ...759L..27P,2021ApJ...911..128K}.
Adopting values from 
% EKO22
\citet{2022ApJ...929L..18E},
\begin{equation}\label{koeqn}
\epsilon_{\rm ff}\rm{(EKO)} = 0.30\, \exp(-2.018\, \alphavir^{1/2}),
\end{equation} 
so \tdep\ should increase exponentially with $\alphavir^{-0.5}$ because $\tdep \propto 1/\epsff$.

The fact is that none of these expected relations are seen in the data, with the exception of the $\tdep - \sigmav$ relation in the \mw, and we noted earlier that this relation is quite sensitive to a few data points. We will explore possible explanations of each relation after further consideration of the effect of spatial resolution on the volume density and thus free-fall time. 

\subsection{Effect of Resolution Redux}\label{sec:reseffect}

The similarity of the average volume density (\nbar) across samples and tracers (Table \ref{tab:tracer} and Figure \ref{fig:surfdens}) is quite remarkable. The average values of the log of \nbar, translated to actual values are essentially all 10-30 \cmv. This density is much lower than usually assumed for molecular clouds. The average densities are far below the critical density, which is often erroneously attributed to any emission from a molecule, but more consistent with effective densities, which will produce a 1 \kkms\ emission line for an extinction of 1 mag 
%Evans 99, Shirley
\citep{1999ARA&A..37..311E,2015PASP..127..299S}. 
The effective densities  are 15~\cmv\ for CO \jj10\ and 60~\cmv\ for \jj21\ \citep{2021ApJ...920..126E}. The mean densities of the clouds are smaller than the effective densities, but not greatly so. 

However, molecular gas with a density of 10 \cmv\ and the usually assumed $\tk = 10$ K would have very low pressure.  To explore this issue further, we utilize the wide range of distances present in the 
%MD17
\citetalias{MD17} catalog to study how \nbar\ varies with distance and thus effective spatial resolution on scales smaller than 150 pc. 
%MD17
\citetalias{MD17} already discussed the effects of finite resolution on the radius, mass, and virial parameter in their Appendix C. Because at least $N_{\rm min} = 5$ pixels were required to define a cloud, the effective angular resolution is larger than the beam size by $\sqrt {N_{\rm min}}$. We discuss the effects on parameters of interest in terms of this effective spatial resolution given by the effective beam size times the distance. To avoid including the clouds in the CMZ, we limited the sample to $\rgal \geq 0.5$ kpc.

\begin{figure}[ht!]
\includegraphics[width= 0.47\textwidth]{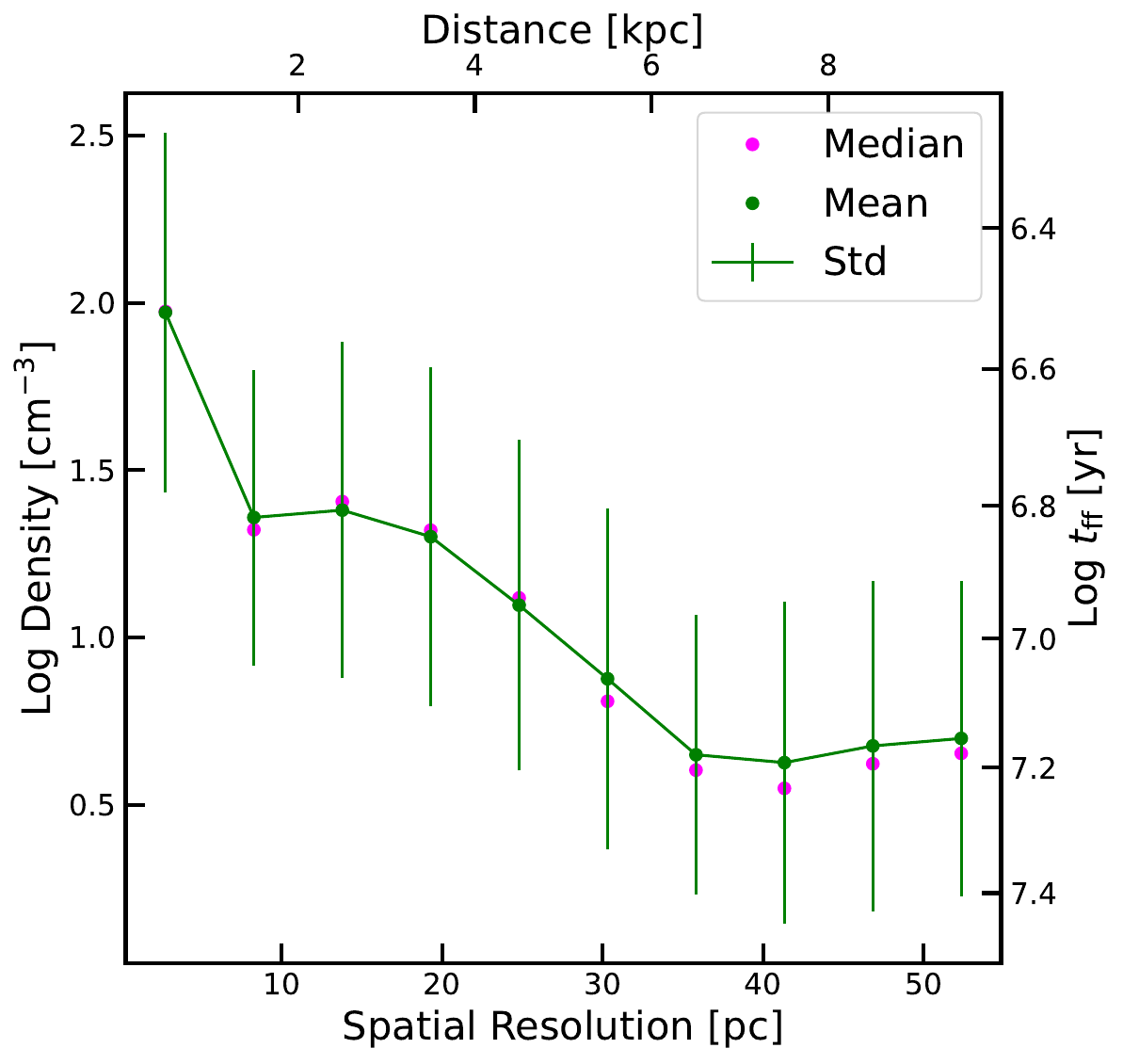}
\caption{The median, mean, and standard deviation of $\log \nbar$ in bins of 1 kpc of distance from the Sun (upper axis). The effective resolution is indicated on the lower axis. The right axis shows the corresponding values of $\log \tff$.
}
\label{fig:nbarvsd}
\end{figure}

\begin{figure}[ht!]
\includegraphics[width= 0.47\textwidth]{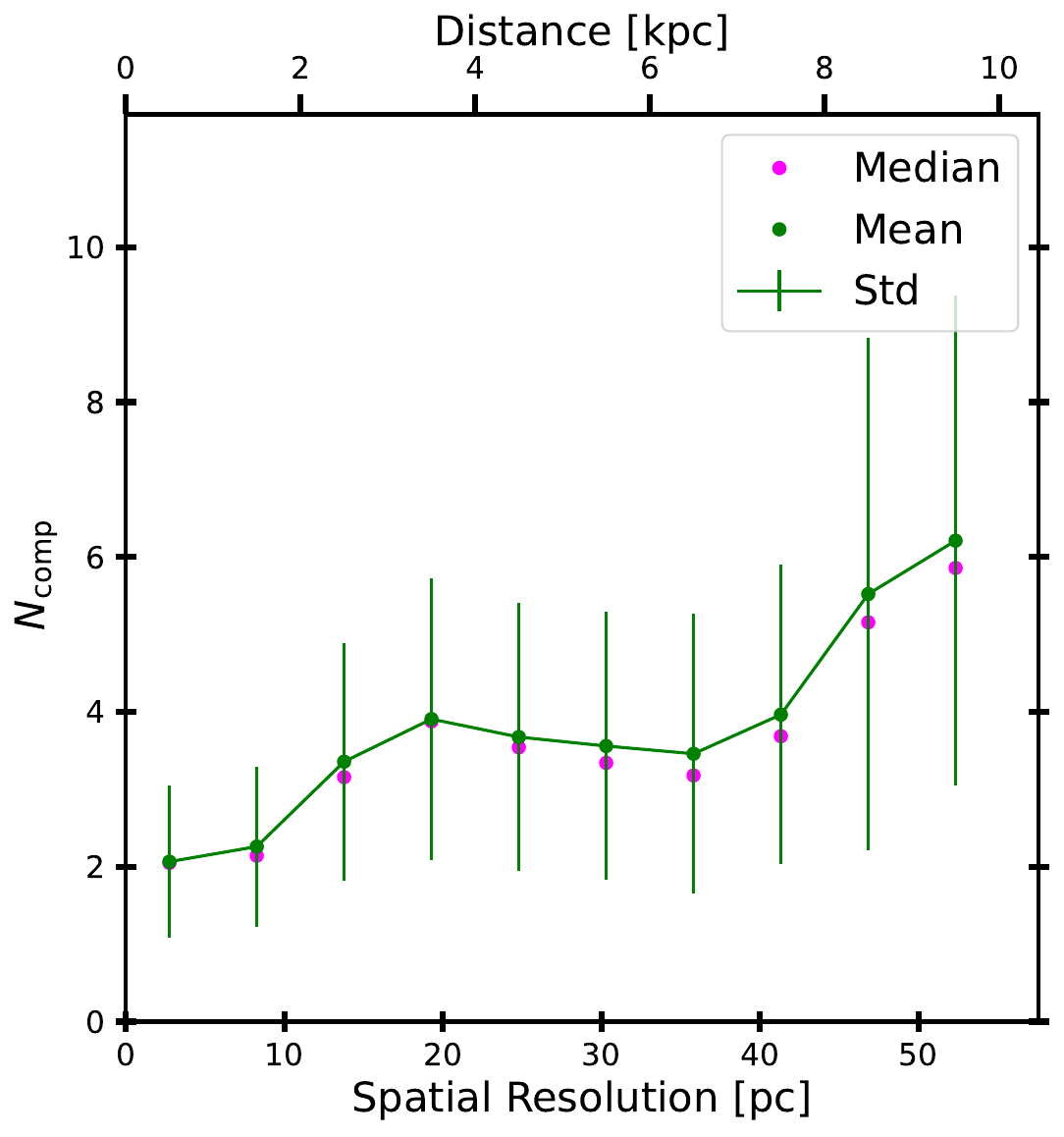}
\caption{The median, mean, and standard deviation of the number of Gaussian components per cataloged cloud in 
%MD17
\citetalias{MD17} in bins of 1 kpc of distance from the Sun (upper axis). The effective resolution is indicated on the lower axis.
}
\label{fig:ncompvsd}
\end{figure}

Figure \ref{fig:nbarvsd} shows the density versus resolution at distances from 0 to 10 kpc. The corresponding \tff\ from equation \ref{eq:tff} is indicated on the right. While the median and mean log(\nbar) over the whole sample correspond to about $\nbar = 10$ \cmv, this number clearly varies substantially with distance (resolution). While clouds within 1 kpc have mean \nbar\ close to 100 \cmv, the estimated \nbar\ drops with distance. We interpret this to mean that many clouds are unresolved and the fraction of empty space within the identified cloud grows with distance. Beyond 35 pc resolution, the mean \nbar\ levels off, suggesting that the observations are essentially in the cloud counting regime, and larger resolutions add clouds just as fast as individual clouds are beam-diluted. To test this idea, we plot the number of Gaussian components combined to define a cloud by 
%MD17
\citetalias{MD17}. In other definition schemes, these might be clouds being combined into a complex. The number is 2 for nearby clouds, rises to a plateau around 4 for 3-6 kpc, then starts rising, reaching 6 at 9 kpc (Figure \ref{fig:ncompvsd}).

This effect explains why the mean \nbar\ over the whole sample is so low and why it agrees with the \citetalias{2025ApJ...985...14L} results: for most of the \mw\ and all of the PHANGS galaxies, the average density refers to a region with many smaller clouds and substantial empty space. 
By empty space, we mean gas not emitting in the CO \jj10\ line. This could be CO-dark gas, Cold Neutral Medium (CNM),  Warm Neutral Medium (WNM), or even ionized gas. 

This pattern persists to smaller scales. The maps with the highest spatial dynamic range are those of nearby clouds, like the Taurus molecular cloud, with 3\ee6 pixels in CO 
\citep[e.g.,][]{2008ApJ...680..428G}. 
They confirm that ``clouds" are wispy, highly inhomogeneous, and likely associated with neutral atomic gas. Interestingly, 
%Soler
\citet{2023A&A...675A.206S} calculate a mean volume density (\nbar) of about 10 \cmv\ (after converting from density of nuclei), for the Taurus cloud using dust extinction. The volume density measured on smaller scales within the Taurus cloud varies from 75 \cmv\ to 1200 \cmv\ as one goes from the weakest CO emission to the strongest
% Goldsmith 2008
\citep{2008ApJ...680..428G}.

The nearby clouds are not representative of the more massive regions of star formation, but studies of those also find highly filamentary and fragmented structures on scales much smaller than available to either 
%MD17
\citetalias{MD17} or PHANGS
%  Sanhueza19 (ASHES), Xu23 (ASSEMBLE), Schisano (ALMAGAL), 
\citep[e.g.,][and references therein]{2019ApJ...886..102S,2024ApJS..270....9X, 2026A&A...707A.221S,2020ApJ...894..103E,2025ApJ...983..133P,2025ApJ...993..193O}.
Such a ``Swiss cheese with many holes and little cheese'' model was originally suggested by
%Zuckerman and Evans
\citet{1974ApJ...192L.149Z}.

\begin{figure}[ht!]
\includegraphics[width= 0.47\textwidth]{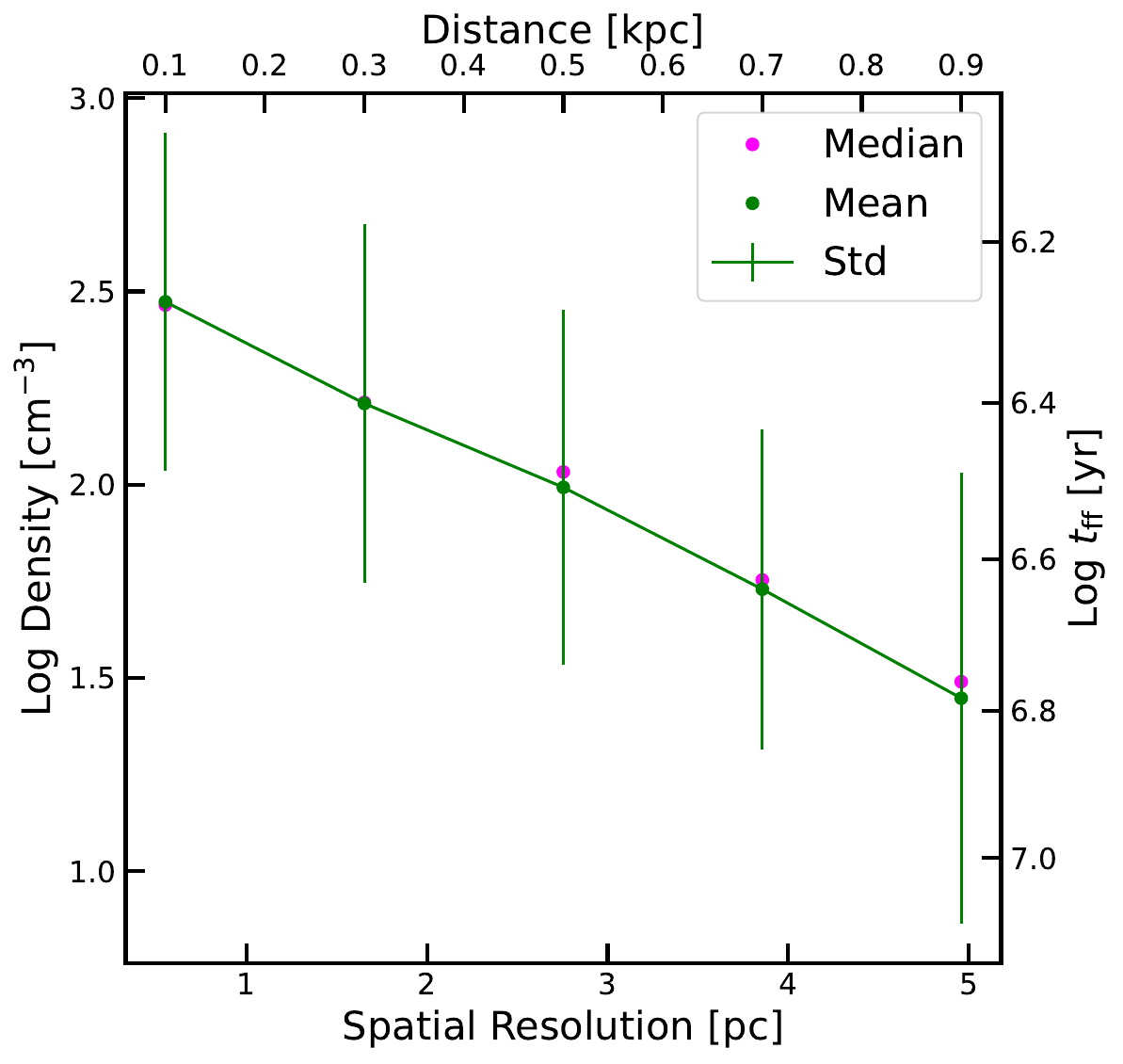}
\caption{The median, mean, and standard deviation of $\log \nbar$ in bins of 0.2 kpc of distance from the Sun (upper axis). The effective resolution is indicated on the lower axis. The right axis shows the corresponding values of $\log \tff$.
}
\label{fig:nbarvsdbu}
\end{figure}

One might hope to use the nearby clouds in the 
%MD17
\citetalias{MD17} sample to define the actual \nbar\ to use for calculating free-fall times of unresolved clouds. However, a blow up of the closest 1 kpc shows that \nbar\ continues to rise with smaller resolution, reaching 300 \cmv\ for the bin from 0 to 0.2 kpc (Figure \ref{fig:nbarvsdbu}).
The density derived from observations of a cloud is always a function of the resolution and the tracer observed.

\begin{figure*}[ht!]
\includegraphics[width= 0.50\textwidth]{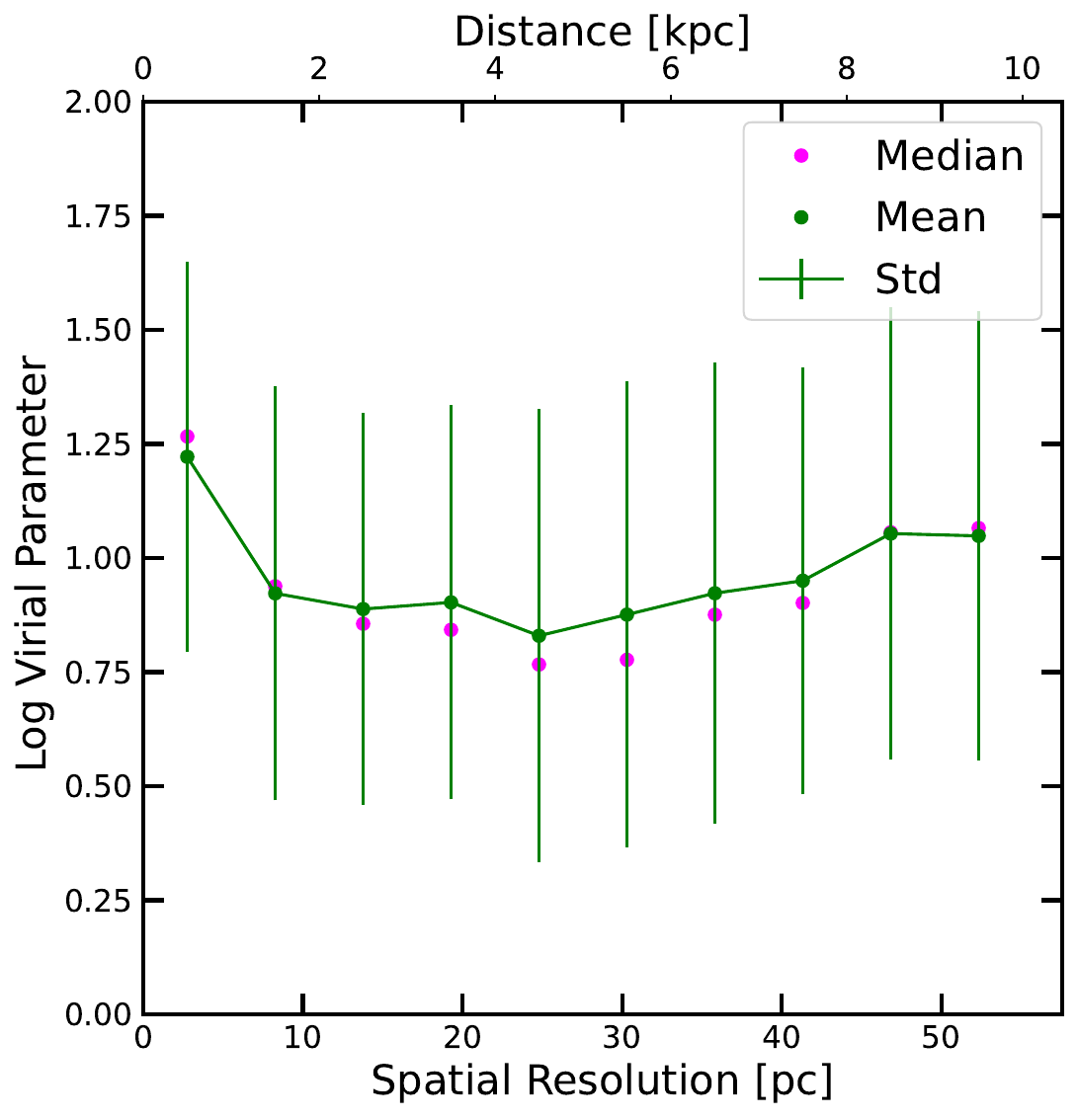}
\includegraphics[width= 0.50\textwidth]{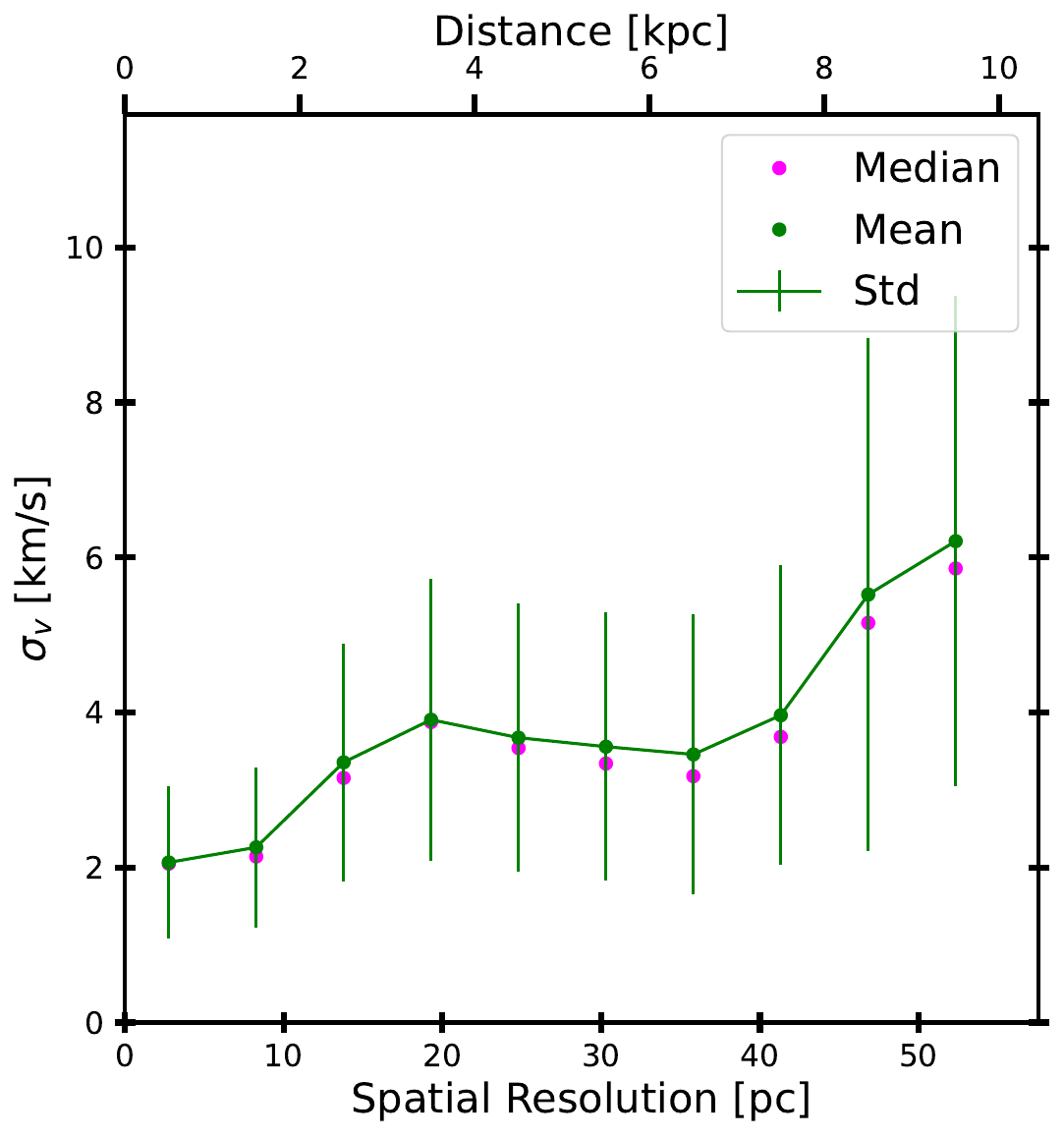}
\caption{The median, mean, and standard deviation of \alphavir\ and \sigmav\ in bins of 1 kpc of distance from the Sun (upper axis). The effective resolution is indicated on the lower axis.
}
\label{fig:avirvsd}
\end{figure*}

The virial parameter shows little change with resolution from 1-50 pc (Figure \ref{fig:avirvsd}).
Similarly,
%Oakes 25
\citet{2025ApJ...993..193O} find no emergent scale for bound objects between 3 and 300 pc in an analysis of molecular structure in NGC253.
In the \mw, clearly bound objects emerge at scales of about 1 pc, where they depart from Larson's laws 
% Evans Plume, Peretto
\citep{1997ApJ...476..730P,2023MNRAS.525.2935P,2021ApJ...920..126E}. 
Infall is well established at about 0.1 pc 
% Evans B335 Yang20 BHR71
\citep{2015ApJ...814...22E, 2020ApJ...891...61Y}.
These two scales are associated with the terms dense clumps and protostellar cores, respectively
%  Bergin and Tafala, McKee and Ostriker
\citep{2007ARA&A..45..339B,2007ARA&A..45..565M}. 
The velocity dispersion is also relatively insensitive to distance (resolution) with mean values around $4 \pm 2$ \kms. There is a rise at about 8 kpc to values closer to 6 \kms (Figure \ref{fig:avirvsd}).

\subsection{Observations versus Expectations }\label{sec:obsvtheory}

The fact that almost no clouds in the PHANGS samples or in the 
%MD17
\citetalias{MD17} catalog are fully resolved could help explain the lack of correlation of \tdep\ and \tff. The value of \tff\ calculated from a collection of clouds will be overestimated and not really relevant to the time scale for depletion due to star formation. 
Highly structured clouds with low average densities can also explain the fact that \tdep\ depends weakly on surface density (Figure \ref{fig:ALf4-tff-MW-alt} and Table \ref{tab:correlations2}).

The \citetalias{2025ApJ...985...14L} data show a negative correlation between \tff\ and \sigmam,
while the \mw\ data show no correlation at all.  The majority of the data points for both the PHANGS galaxies and the \mw\ lie at $\sigmam < 120$ \msunpc, where no strong correlation exists between surface density and SFR in \mw\ clouds
%Heiderman 2010 and Lada 2010, 2012
\citep[e.g.][]{2010ApJ...723.1019H,2010ApJ...724..687L,2012ApJ...745..190L}.
If the \citetalias{2025ApJ...985...14L} data are restricted to the (only 59) points with $\sigmam > 120$ \msunpc, the values of $a$ (Equation \ref{eq:tdep-sigm}) become 0.63 or 0.77 for the normalized data. The data points that meet that criterion are mostly near galaxy centers, with median galactocentric radius of 1.5 kpc.

If the observations are primarily in the cloud-counting regime, and star formation is primarily dependent on the total mass of molecular gas, a simple prediction is that $ \sigmasfr \propto \sigmam$, and thus $\tdep \propto \sigmam/\sigmasfr$ is constant with \sigmam, as seen in the \mw\ data.
Relations like KS work well on full galaxy scales and in some cases on structures within molecular clouds, but not when averaged over regions with a mix of star-forming gas and more diffuse gas. In addition, the measures of star formation rate are often displaced in time and space from the molecular gas, as 
first noted by 
% Schruba 10
\citet{2010ApJ...722.1699S}, and emphasized by
% Kruijssen14
\citet{2014MNRAS.439.3239K}.

The lack of correlation between \tdep\ and \alphavir\  suggests that virial parameters on 150 pc scales are not useful predictors of star formation rate. 
In contrast, including virial parameters on a cloud-by-cloud basis in the 
%MD17
\citetalias{MD17} catalog produced good agreement on the overall star formation rate of the \mw, as well as the radial distribution
% EKO
\citep{2022ApJ...929L..18E}.
\citetalias{2025ApJ...985...14L} suggest that the high velocity dispersion, high virial parameter regions are in the inner regions of galaxies, where higher densities and contributions from stellar surface density overcome the higher dispersions. This idea is supported by the trend to higher surface densities and shorter \tff\ in the very central regions of both PHANGS galaxies and the \mw\ in Figure \ref{fig:ALf8-MW} and by Figure \ref{fig:ALf6-sigv-MW-alt} showing that the hexagons with high \avirhex\ are in the inner \mw.

The one predicted relation that does appear in the data is that $\tdep \propto \sigmav^{-1}$. The fit to the PHANGS data finds an exponent of about $-0.5$ but the MW data find $-1.14 \pm 0.28$, reasonably consistent with the simple prediction. (Recall, however, that we simply averaged cloud velocity dispersions to construct the cell-level value, as noted in \S \ref{sec.proc}.) In the crossing time picture, the velocity dispersion over 150 pc reflects the time for structural change: clouds merging, moving in or out of the cell, etc. In the Keto-Heyer picture, it would reflect more of an equilibrium state, but only if $\alphavir$ is fixed. In fact, \alphavir\ varies widely in either the 
%MD17
\citetalias{MD17} or 
%Rice
\citetalias{2016ApJ...822...52R} cloud catalogs.

\subsection{Relation to Discussion in \citetalias{2025ApJ...985...14L}}\label{sec:L25disc}

A substantial discussion of how to interpret the PHANGS results is found in \S 4 of
\citetalias{2025ApJ...985...14L}. All the interpretation in the previous section should be seen as ``in addition to'', rather than as ``instead of'', that discussion. The fact that the individual locations of star formation are not fully resolved even in the \mw\ adds emphasis to the points made there. The failure of expected relations to emerge in the data on kpc scales applies to the \mw\ as well. We agree with \citetalias{2025ApJ...985...14L}, who argue that the utility of these measurements lies in comparison to simulations that predict observables on similar scales. The large range of distances, and thus spatial resolution, available in the \mw\ data allows more stringent tests of simulations.

The \mw\ adds assurance that the trends seen in the PHANGS galaxies are not an artifact of observing at a large distance. We agree with the conclusion in \citetalias{2025ApJ...985...14L} that ``there is no clear evidence for a specific scale where the surface density traces physical density," and Figures \ref{fig:nbarvsd} and \ref{fig:nbarvsdbu} extend that conclusion to scales as small as 1 pc.

\section{Summary}\label{sec:summary}

Analysis of cloud catalog data for the \mw\ with a method similar to that used for PHANGS galaxies produces broadly similar results. If the depletion time is plotted versus other parameters for which a relationship might be expected, the result is a very large scatter and relatively weak relationships. Simulations of star formation in galaxies should be able to reproduce these observations. The data in the \mw\ can be used to test simulations on still smaller scales by using the range of distances, hence resolutions, available.

Many of the trends that do exist and the differences between the \mw\ and the PHANGS galaxies can be traced to different conditions that exist near the centers of galaxies, including the \mw. The radial distributions of many variables in the \mw\ are similar to those found in the PHANGS galaxies, with some differences, perhaps related to the effects of the bar in the \mw\ 
% Evans 26, Baba 26
\citep{2026ApJ...998L..23E,2026PASJ...78.1093B}. More specific results are listed next.

\begin{enumerate}
\item The formula used by \citetalias{2025ApJ...985...14L} for how the luminosity to mass conversion factor (\alphaco) depends on metallicity ($Z$) differs substantially from that used in studies in the \mw.
However, the formula for how $Z$ depends on galactocentric radius also differs substantially from the measured dependence in the \mw. The two effects cancel nearly exactly, so our analysis does not depend strongly on which we use.  This cancellation is fortuitous but probably masks deeper issues. See \S \ref{sec:alphaco} and Figure \ref{fig:alphaz} in particular.

\item The coarse spatial resolution of the \citetalias{2025ApJ...985...14L} data set (150 pc) and the surface density limitation ($\sigmam \geq 4$ \msunpc) mean that observers from another galaxy would miss many clouds in the \mw\ catalog and underestimate the surface density of those they detect. However the mass-weighting scheme when averaging into hexagons of \citetalias{2025ApJ...985...14L} increases the resulting measures of surface density to be more similar to the underlying cloud level distribution. See \S \ref{sec:resolution} and Figure \ref{fig:res1}.

\item The sensitivity limit ($\ico \geq 0.6$ \kkms) of the \citetalias{2025ApJ...985...14L} analysis has a minimal effect on the inferred properties. See \S \ref{sec:sensitivity}. Better sensitivity in surveys like PHANGS would have few benefits.

\item The tracer used (CO \jj10, CO \jj21, extinction) matters for the inferred surface density and average volume density, but only at the factor of two level. See \S \ref{sec:tracers} and Figure \ref{fig:surfdens}.

\item The method of cloud definition 
%MD17
(\citetalias{MD17} versus 
%R16
\citetalias{2016ApJ...822...52R}) affects measures of the virial parameter substantially. See \S \ref{sec:tracers} and panel c of Figure \ref{fig:surfdens}.

\item The relations between \tdep\ and other variables expected from simple considerations (\S \ref{sec:theory}) are not generally seen, neither in the \citetalias{2025ApJ...985...14L} data nor in the \mw. There are a number of explanations for this, as detailed in \citetalias{2025ApJ...985...14L} and \S \ref{sec:obsvtheory}. Two of the more important explanations are that gas and star formation behave differently near galactic centers and that the simple considerations do not account for hierarchical cloud structure (\S \ref{sec:reseffect}).

\item The method of averaging free-fall times has a moderate effect. Mass-weighting the speeds ($1/\tff$) as adopted by \citetalias{2025ApJ...985...14L} produces an effective free-fall time about 30\% less than averaging \tff. See Appendix \ref{sec:tffdef}.

\item Averaging virial parameters because of coarse resolution causes formulae using \alphavir\ to underestimate the efficiency per free-fall time and thus star formation rate, though mass-weighting mitigates this problem. See Appendix \ref{sec:aviravg}.

\end{enumerate}

\begin{acknowledgments}

We are grateful to C. McKee and D. Hollenbach for interesting comments on the mean cloud densities in Table \ref{tab:tracer}.
This research used the Canadian Advanced Network For Astronomy Research (CANFAR) operated in partnership by the Canadian Astronomy Data Centre and The Digital Research Alliance of Canada with support from the National Research Council of Canada the Canadian Space Agency, CANARIE and the Canadian Foundation for Innovation. TThis paper makes use of the following ALMA data,
which have been processed as part of the PHANGS–
ALMA CO (2-1) survey:
ADS/JAO.ALMA\#2012.1.00650.S,
ADS/JAO.ALMA\#2013.1.00803.S,
ADS/JAO.ALMA\#2013.1.01161.S,
ADS/JAO.ALMA\#2015.1.00121.S,
ADS/JAO.ALMA\#2015.1.00782.S,
ADS/JAO.ALMA\#2015.1.00925.S,
ADS/JAO.ALMA\#2015.1.00956.S,
ADS/JAO.ALMA\#2016.1.00386.S,
ADS/JAO.ALMA\#2017.1.00392.S,
ADS/JAO.ALMA\#2017.1.00766.S,
ADS/JAO.ALMA\#2017.1.00886.L,
ADS/JAO.ALMA\#2018.1.01321.S,
ADS/JAO.ALMA\#2018.1.01651.S,
ADS/JAO.ALMA\#2018.A.00062.S,
ADS/JAO.ALMA\#2019.1.01235.S,
ADS/JAO.ALMA\#2019.2.00129.S. ALMA is a partnership of ESO (representing
its member states), NSF (USA) and NINS (Japan), together with NRC (Canada), MOST and ASIAA (Taiwan), and KASI (Republic
of Korea), in cooperation with the Republic of Chile. The Joint ALMA Observatory is operated by ESO, AUI/NRAO and NAOJ.
NJE is grateful to the Astronomy Department of the University of
Texas for research support, INAF-IAPS (Roma, Italy) for
hospitality during a visit, and to H. Khan-Farooki, along with two
anonymous cornea donors, for the gift of sight.
\end{acknowledgments}

\software{astropy
\citep{2013A&A...558A..33A,2018AJ....156..123A,2022ApJ...935..167A}, scipy \citep{2020SciPy-NMeth}
}

\appendix

%\section{Appendix information}

\section{Some Details of comparisons}\label{app:details}

We note here some relevant facts about the comparison of the \mw\ survey and the PHANGS-ALMA survey.
The threshold value for PHANGS-ALMA is 0.6~\kkms\ \citepalias{2025ApJ...985...14L}. 
Using $R_{21} = 0.65$ 
%Schinnerer and Leroy 24
\citep{2024ARA&A..62..369S}, this corresponds to 1.0~\kkms\ for the \jj10\ line. For the
%Dame et al 01 
\citet{Dame01}
survey, the sensitivity varies but is always less than 0.25 K (Fig. 22 in 
%MD17
\citetalias{MD17}).
With a velocity resolution of 1.3 \kms, this corresponds to 0.33 \kkms. 
Thus the sensitivity of the \mw\ survey is at least 3 times better than the threshold for PHANGS-ALMA.
For the great majority of the data, the intensity sensitivity is less than 0.13 \kkms, 8 times better than the PHANGS-ALMA threshold. We consider the effects of this difference in \S \ref{sec:sensitivity}.

The spatial resolution of PHANGS-ALMA depends on the distance to the galaxy, but they have been convolved to a common spatial scale of 150 pc for the primary analysis. Some analysis with higher resolution (as small as 60 pc for 9 galaxies) for the more nearby galaxies was also done. For the 8\farcm5 beam of the Dame survey, the resolution  varies dramatically with heliocentric distance. At 20 kpc, it is 50 pc, compared to 0.3 pc for the closest molecular clouds.  Especially with subsets located closer to the Sun, we can examine the effects of spatial resolution (\S \ref{sec:resolution} and \S \ref{sec:reseffect}).

For the \mw\ clouds, we use $\sigmamcloud = \masscloud/\areacloud$, where \areacloud\ is the area of the cloud; for average density, we use $\nbarcloud = \masscloud/V_{\rm cloud}$, where $V_{\rm cloud}$ is the volume of the ellipsoid that defines the cloud in 
%MD17
\citetalias{MD17}. For the \mw, our cells are squares of 150 pc on a side. To compute the volume, we assume a scale height of 100~pc for consistency with \citetalias{2025ApJ...985...14L}. Then the smoothed surface density of a cell (\sigmamcell) is the mass of the cell over the area of the cell. The smoothed volume density (\nbarcell) is the mass of the cell divided by the volume of the cell. For the hexagons, we use the same formulae for PHANGS and the \mw, mass-weighting the averages of values in the cells.

\citetalias{2025ApJ...985...14L} define \sigvp\ as the effective (1D) velocity dispersion within their 150 pc resolution, corrected for galaxy inclination by multiplying by $(\cos i)^{0.5}$. 
%Heyer01
\citet{2001ApJ...551..852H}
 defines ``effective dispersion" in Equation A7 as $\sigma_v = \delta v/( 8 \ln 2)^{0.5}$, where $\delta v$ is the FWHM of summed spectrum of all pixels. 
Because we work with catalogs, we simply average the 
%MD17
\citetalias{MD17} catalog values of velocity dispersions (\sigvcloud) for all clouds in a cell to get \sigvcell. As noted above, this may underestimate the full dispersion measured by \citetalias{2025ApJ...985...14L}, but it more accurately describes the actual cloud-level dispersion.

\citetalias{2025ApJ...985...14L} define $\avirp = 5 \sigvp^2 R_{\rm pix}/(f G \massp)$;  $f$ is a factor that depends in the cloud density structure
%Bertoldi and McKee
\citep{1992ApJ...395..140B}, but it was assumed to be unity in \citetalias{2025ApJ...985...14L} and we assume the same.
We define $\avircell = 5 \sigvcell^2 R/(G \masscell)$.
\citetalias{2025ApJ...985...14L} use $R_{\rm pix} = (l^2H/8 \cos i)^{1/3}$, where $l = 150$ pc and $H = 100$ pc. For $\cos i = 1$, this yields 65.5 pc.
Our \avircell\ is computed from the same formula, with $R = 65.5$ pc, the averaged velocity dispersion (\sigvcell), and the total mass in a cell (\masscell). For the hexagons, we calculate a mass-weighted average of \avircell, consistent with \citetalias{2025ApJ...985...14L}.

For all quantities labeled with $\mean{x}_{\rm hex}$, the values are mass-weighted averages over the values in cells:
\begin{equation}\label{eq:massweight}
\mean{x}_{\rm hex} = \frac{\sum_i M_i x_i}{\sum_i M_i}
\end{equation}

\section{Stellar Surface Density}\label{sec:starsurfden}

As noted in \S \ref{sec:alphaco}, the CO luminosity to mass conversion factor can depend on stellar surface density through the starburst factor
%Schinnerer and Leroy
\citep{2024ARA&A..62..369S}.
This becomes important only where $\sigmastar > 100$ \msunpc. 
The standard exponential disk model for stellar density 
% Binney 15 
\citep{2015MNRAS.454.3653B}
does not predict values of \sigmastar\ above this value, but 
% Lian 24 Nature Ast
\citet{2024NatAs...8.1302L}
have used APOGEE and GAIA data to derive nonparametric \sigmastar\ distributions of bolometric luminosity in age bins from \rgal\ = 0 to 17 kpc. Their Figure 1 shows the total luminosity surface density versus \rgal.  Assuming mass-to-light of unity, it exceeds 100 \msunpc\ only in the innermost ring, $\rgal = 0$ to 2 kpc. Much of that is due to very old stars which have large heights and may not contribute to the gravity acting on the molecular gas, as shown by the data including only ages less than 8 Gyr. The data for their youngest bin (0-2 Gyr) shows a very strong dip, similar to that seen in the molecular and star formation distributions.

The starburst factor in the \mw\ would be unity everywhere but perhaps in the central 2 kpc. If we use the 
%Lian 24
\citet{2024NatAs...8.1302L}
data, assume mass-to-light of unity, and
take the implied total stellar density of $\sigmastar = 1.2\ee3$ \msunpc, the starburst factor would be 0.53. If we use only the stars younger than 8 Gyr, the factor would be 0.74. We use the latter.

The stellar distribution also affects simulation of the extragalactic form for \alphaco(\rgal) through the method of determining $Z(\rgal)$, which uses a form that depends on $\rgal/\reff$, where \reff\ is the effective
radius, containing half the light of the galaxy. To simulate this for the \mw, we used $\reff = 4.75$ kpc
% Imig 25
\citep{2025ApJ...990..203I}.

\section{Definitions of \tff}\label{sec:tffdef}

The definition of \tff\ in Equation \ref{eq:tffL25} emphasizes the regions most active in star formation by mass-weighting and averaging speeds, rather than \tff. To test the effects, we compared results with two other averaging methods: a straight average of free-fall times, and an unweighted average of speeds (1/\tff). The test was performed in the step of averaging cells into hexagons for the \mw\ data, using the \citetalias{2025ApJ...985...14L} criteria. After computing the three quantities for each hexagon, we computed the mean values over all hexagons for each of the three methods. The straight average of \tff\ leads to the longest average (11.1 Myr); the average of speeds yields 9.2 Myr; and the mass-weighted average of speeds is 7.4 Myr. Using mass-weighted speeds does not have a major effect, but it decreases the timescale by about 30\%. Because of the non-linearity, it is certainly better to average speeds than free-fall times.

\section{Effects of averaging \alphavir}\label{sec:aviravg}

The virial parameter is also mass-weighted in the L25 method. To test the effect of this, we computed it with and without mass-weighting, using the same method described in Appendix \ref{sec:tffdef}. The mean value with mass weighting is 44\% smaller than that without mass weighting.

Another question is whether averaging \alphavir\ misleads us about star formation. To test this, we calculated the star formation rate of the \mw\ three ways, all using the EKO formula (Equation \ref{koeqn}). First, we redid the EKO method cloud by cloud with the various cuts above as a check and got $\sfr = 1.39$ \msunyr, only slightly lower than the 1.47 \msunyr\ in 
%E25
\citet{2025ApJ...980..216E}. 
Second, we calculated with the total mass, the average \epsff\ using each hexagon's mass-weighted \alphavir, and the mass-weighted inverse free-fall time, with the result of 1.06 \msunyr, two-thirds the correct value. 
Third, we computed with the total mass, the \epsff\ calculated from the mean mass-weighted \alphavir, and the mass-weighted inverse free-fall time, using the values in Table \ref{tab:statskey}, with the result of 0.40 \msunyr, about one-third the correct value. 

One needs to average \epsff, not \alphavir\ because of the non-linearity. The most bound clouds have a high weight in the star formation process. For that reason, any averaging (or lower spatial resolution) will decrease the predicted star formation rate  because the active clouds are diluted by the inactive clouds.

On the other hand, the mass-weighting has decreased \alphavir\ on average (compare values in Table \ref{tab:res} and \ref{tab:statskey}). Many of the clouds with high \alphavir\ in the 
%MD17
\citetalias{MD17} catalog have low mass and are down-weighted in the L25 method. The value of \avirhex\ in Table \ref{tab:statskey} is however still larger than that for the PHANGS sample. The 
%MD17
\citetalias{MD17} method of cloud identification also produces values of \alphavir\ that are larger than those in a less complete catalog of \mw\ clouds compiled by 
%Rice
\citetalias{2016ApJ...822...52R}, as can be seen in the entries for each in Table \ref{tab:tracer}.

\section{Effect of Assumptions about \alphaco}\label{app:alphaco}

\citetalias{2025ApJ...985...14L} noted that the choice of assumptions about \alphaco\ had a substantial impact on their results. We confirm this fact using the data in their table. Making the same assumption of a constant $\alphaco = 4.35$, we find weaker correlations in general in both the PHANGS and \mw\ data, as shown in Table \ref{tab:corr2fixeda}. The differences in the fits for the \mw\ and the PHANGS galaxies are even less, except for the velocity dispersion, on which \tdep\ still depends more strongly in the \mw. 

%----------Table correlations with Uncertainties-----------------------   
\begin{deluxetable*}{l l r r r r r r r} 
\tablecaption{Correlations and Fits for fixed \alphaco \label{tab:corr2fixeda}} 
\tablewidth{0pt} 
\tablehead{\colhead{Sample} & \colhead{$x$} &  \colhead{$r_{\rm s}$} & \colhead{$p$} & \colhead{$m$}  & \colhead{$\sigma_{\rm m}$} & \colhead{$x_0$}   & \colhead{$b$}   & \colhead{$\sigma_{\rm b}$} }  
\startdata 
\hline 
PHANGS & log \tffhex & $-0.18$ & 3.11e-07 & $-0.22$ & 0.05 & 6.91 & 9.37 & 9.95e-03  \\ 
MW  & log \tffhex & $-0.00$ & 9.57e-01 & $0.04$ & 0.25 & 6.95 & 9.25 & 3.67e-02  \\ 
MW (PHANGS) & log \tffhex & $-0.00$ & 9.57e-01 & $-0.06$ & 0.23 & 6.95 & 9.25 & 3.67e-02  \\ 
PHANGS  & log \sigsmhex & $0.13$ & 1.28e-04 & $0.09$ & 0.03 & 1.66 & 9.37 & 1.00e-02  \\ 
MW  & log \sigsmhex & $0.01$ & 9.31e-01 & $-0.00$ & 0.13 & 1.67 & 9.25 & 3.68e-02  \\ 
MW (PHANGS) & log \sigsmhex & $0.01$ & 9.31e-01 & $0.04$ & 0.12 & 1.67 & 9.25 & 3.68e-02  \\ 
PHANGS  & log \sigvcellhex  & $0.14$ & 7.32e-05 & $0.21$ & 0.05 & 0.74 & 9.37 & 1.02e-02  \\ 
MW  & log \sigvcellhex  & $-0.18$ & 2.64e-02 & $-0.72$ & 0.26 & 0.77 & 9.24 & 3.59e-02  \\ 
MW (PHANGS) & log \sigvcellhex  & $-0.18$ & 2.64e-02 & $-0.47$ & 0.27 & 0.77 & 9.24 & 3.59e-02  \\ 
PHANGS  & log \avirhex & $-0.02$ & 5.36e-01 & $-0.00$ & 0.05 & 0.26 & 9.38 & 4.58e-02  \\ 
MW  & log \avirhex & $-0.11$ & 1.57e-01 & $-0.19$ & 0.14 & 0.57 & 9.25 & 3.64e-02  \\ 
MW (PHANGS) & log \avirhex & $-0.11$ & 1.57e-01 & $-0.23$ & 0.13 & 0.57 & 9.25 & 3.64e-02  \\ 
\enddata  
\tablecomments{ 1. MW (PHANGS) denotes MW data restricted to range of PHANGS data. \\  } 
 \end{deluxetable*} 

\onecolumngrid  % <--- ADD THIS HERE to force one-column mode again

\begin{figure}[ht!]
\includegraphics[width= 0.47\textwidth]{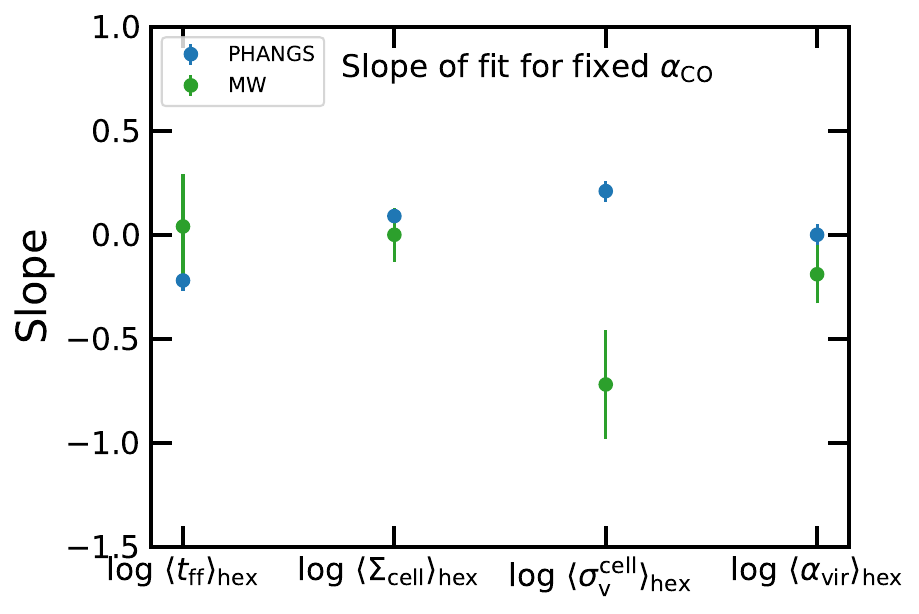}
\caption{The slopes of the fits in Table \ref{tab:corr2fixeda} are plotted with uncertainties for both the PHANGS galaxies and the \mw, with the assumption of fixed $\alphaco = 4.35$.
}
\label{fig:compareslopesfix}
\end{figure}

%\onecolumngrid  % <--- ADD THIS HERE to force one-column mode again

\bibliography{cite}{}
\bibliographystyle{aasjournal}

\end{document}